\documentstyle[11pt,aaspp4,flushrt]{article}
\lefthead{Diaferio et al.}
\righthead{Mock Redshift Surveys}
\begin{document}
\title{CLUSTERING OF GALAXIES IN A HIERARCHICAL UNIVERSE: \\ III. 
MOCK REDSHIFT SURVEYS}
\author{Antonaldo Diaferio, Guinevere Kauffmann, J\"org M. Colberg, 
\& Simon D. M. White\footnote{diaferio,gamk,jgc,swhite@mpa-garching.mpg.de}}
\affil{Max-Planck Institut f\"ur Astrophysik, Karl-Schwarzschild-Str. 1,
D-85740 Garching, Germany}
\authoraddr{address}

\begin{abstract}

This is the third paper in a series which combines 
$N$-body simulations and semi-analytic modelling to
provide a fully spatially resolved simulation of the galaxy formation
and clustering processes. Here we extract mock 
redshift surveys from our simulations:
a Cold Dark Matter model with either $\Omega_0=1$ ($\tau$CDM)
or $\Omega_0=0.3$ and $\Lambda=0.7$ ($\Lambda$CDM). We compare 
the mock catalogues with the northern region (CfA2N) of the Center
for Astrophysics (CfA) Redshift Surveys. We study the properties of
galaxy groups and clusters identified using standard observational
techniques and we study the relation of these groups to real virialised systems. 
Most features of CfA2N groups are reproduced quite well by both models with
no obvious dependence on $\Omega_0$.
Redshift space correlations and pairwise velocities are also similar in the two cosmologies.
The luminosity functions predicted by our galaxy formation models depend
sensitively on the treatment of star formation
and feedback. For the particular choices of Paper I they agree poorly
with the CfA survey. 
To isolate the effect of this discrepancy  on our mock redshift surveys,
we modify galaxy luminosities in our simulations to reproduce the CfA
luminosity function exactly.  This adjustment improves 
agreement with the observed abundance of groups, which
depends primarily on the galaxy luminosity density, but 
other statistics, connected more closely with the underlying mass distribution,
remain unaffected.
Regardless of the luminosity function adopted, modest differences 
with observation remain.
These can be attributed to the presence of the ``Great Wall''
in the CfA2N. It is unclear whether the greater coherence of the real structure is
a result of cosmic variance, given the relatively small region studied, or reflects
a physical deficiency of the models.

\end{abstract}


\section{INTRODUCTION}

Groups of galaxies are weak enhancements in the projected galaxy distribution.
Velocity information from redshift surveys 
confirms, however, that they are indeed true spatial density enhancements 
(e.g. \cite{Geller83}; \cite{Tully87}; Ramella, Geller \& Huchra 1989; 
Ramella, Pisani \& Geller 1997), 
although probably still far from equilibrium
(e.g. \cite{Nolthenius87}; \cite{Giuricin88}; \cite{Diaferio93};  
Moore, Frenk \& White 1993; 
\cite{Mamon93}).
Groups probe the intermediate scale between galaxies and clusters (e.g. \cite{Mahdavi98}) 
and can provide important constraints on the amount of dark matter, on the formation and
evolution of galaxies, and on the properties of the intergalactic medium
(e.g. \cite{Geller83}; \cite{Mulchaey96}; \cite{Hickson97}; \cite{Davis98};
\cite{Ponman98}; \cite{Ramella98}).

Further information about the dynamics of structure formation come from
anisotropies of the correlation function in redshift space, which are related to moments
of the pairwise velocity distribution. The first moment $\langle v_{12}\rangle$
enters the time evolution equation for the spatial correlation
function, and has so far been estimated directly only for 
the IRAS 1.2 Jy survey (\cite{Fisher94b}; see also \cite{Juszk99}; \cite{Ferr99}). 
Considerably more attention has been devoted to the second moment $\sigma_{12}$,
which measures the random motions of galaxies and can be used to
constrain $\Omega_0$ via the cosmic virial theorem (\cite{Peebles76};
\cite{Davis83}; \cite{Fisher94b}; \cite{Marzke95}). 
Several authors have shown that $\sigma_{12}$ is sensitive
to the presence of clusters within a survey (\cite{Mo93}; \cite{Zurek94}; \cite{Marzke95}; 
\cite{Somerville97}) and so requires a large survey volume to obtain a
robust and unbiased estimate (\cite{Mo93}; \cite{Zurek94}; \cite{Marzke95};
Somerville et al. 1997a; 
\cite{Jing98}). Many early estimates of
$\sigma_{12}$ were substantially below the values
predicted by the popular CDM model and its variants (see e.g. \cite{Marzke95} and references
therein). Larger values were found by Mo et al. (1993), by Marzke et al.
(1995) for the Center for Astrophysics (CfA, hereafter) surveys, and by Jing et al. (1998) for
the Las Campanas Redshift Survey (LCRS, hereafter). The latter results are for the
largest two surveys currently available and suggest that the large
velocity bias previously advocated to reconcile CDM models
with observations (e.g. \cite{Carlberg91}; \cite{Couchman92}; 
\cite{Gelb94}) may be unnecessary.  

Comparing the observed distribution of galaxies with models of structure formation is problematic.
In particular, if only the dark matter evolution 
is simulated directly, as is usually the case, then some method must be
found to specify the positions, velocities and luminosities of
galaxies before ``mock'' redshift surveys can be constructed 
for direct comparison with real data
(e.g. \cite{Davis85}; \cite{White88}; \cite{Cole98}).
Kauffmann et al. (1998; Paper I hereafter) 
implemented semi-analytic galaxy formation modelling methods on
high resolution $N$-body
simulations of volumes $\sim 10^6h^{-3}$ Mpc$^3$. (Here and below $h$ 
is the Hubble constant $H_0$ in units of  
$100$ km s$^{-1}$ Mpc$^{-1}$.)  This approach enables us to simulate
the physical processes relevant to galaxy formation, evolution and
clustering, and to explore different assumptions about
processes, like star formation and feedback, which occur on scales below the resolution 
limit of our $N$-body simulations.

Standard semi-analytic models (\cite{White91}; \cite{Lacey91}; \cite{Kauffmann93};
\cite{Cole94}; \cite{Baugh96}; \cite{Somerville98}) 
use  Press-Schechter theory (\cite{Press74}) and its extensions (\cite{Bower91};
\cite{Bond91}; \cite{KauffmannW93}; \cite{Lacey93}) to
compute the merging history of dark matter halos. Within these halos, simple
prescriptions, based on observational data and on more detailed simulations, 
regulate gas cooling rates, star formation rates, stellar population
evolution and feedback from stellar winds and supernovae. Dynamical friction
considerations determine the merging rate among galaxies within a
common halo, and so the competition between the formation of spheroids
by merging and the formation of disks through cooling of diffuse gas.
Such models do not, however, provide detailed information about
the clustering and the motions of galaxies.

In our approach (see Paper I), dark matter halos and their evolution
through collapse, accretion and merging are followed directly in the $N$-body simulations.
Other processes are then followed as in standard semi-analytic models.
Diffuse halo gas is assumed to cool and collect in a disk at the
centre of each halo. Stars form in such disks according to simple,
observationally motivated laws. The central galaxy
is identified with the most bound particle in the halo. When
two or more halos merge, the properties of the central galaxy of the 
most massive progenitor are transferred to the most bound particle of the remnant.
Other galaxies from the progenitors become ``satellites''.
These can merge with the central galaxy of the new halo on a
time-scale which is related to their progenitor mass as expected
from numerical experiments (e.g. \cite{Navarro95}). Galaxy properties
such as luminosity, colour, stellar and gas mass, star formation
rate and morphology evolve according to recipes borrowed from earlier semi-analytic work.
We are thus able to analyse clustering as a function of 
galaxy properties and of redshift (Paper I; \cite{KCDW98}, Paper II hereafter). 
 
We investigate two variants
of a Cold Dark Matter (CDM) universe: the $\tau$CDM model, with cosmological
density parameter $\Omega_0=1$, shape parameter 
$\Gamma=0.21$ and Hubble constant $H_0=50$ km s$^{-1}$ Mpc $^{-1}$, and the $\Lambda$CDM model,
with $\Omega_0=0.3$, cosmological constant $\Lambda=0.7$ and $H_0=70$ km s$^{-1}$ Mpc $^{-1}$. 
As discussed in Paper I, in order to obtain an approximate match between models and observations, 
supernova feedback must be very efficient in the  $\tau$CDM model,  
expelling large amounts of the reheated gas and so 
suppressing the formation of galaxies in low-mass field halos. On the 
other hand, feedback must be inefficient in the $\Lambda$CDM models and 
most reheated gas must be retained in order to produce a sufficient 
number of luminous galaxies. Paper I compared the properties of
galaxies in our two models with  a variety of present-day observational data,
among these the luminosity function in the $B$ and $K$ bands, the colour distribution, 
the two-point correlation function, the pairwise velocity dispersion, 
and the mass-to-light ratios of clusters. The free parameters which 
control star-formation and feedback in these models were set by
fitting the I-band Tully-Fisher relation of \cite{Giovanelli97} 
and the mean gas content of a spiral galaxy at a circular velocity of
220 km s$^{-1}$. This 
``normalisation'' choice resulted in fiducial models where certain
other properties, in particular
the luminosity functions, are a relatively poor fit to observation. 
Nevertheless, we used these fiducial models in a follow-up study of the
evolution of clustering to high redshift (Paper II) and, for 
consistency, we will continue to use them
below, even though it turns out that setting
parameters to optimise the luminosity functions gives
better fits to the abundance of groups in magnitude-limited
redshift surveys.

Here we extract wide-angle mock surveys from the simulations
in order to perform a more detailed analysis of small-scale galaxy 
clustering. We compare our models with the northern region of 
the Center for Astrophysics Redshift Survey (CfA2N hereafter; \cite{Geller89};
\cite{Huchra90}; \cite{deLapparent91}; \cite{Huchra95}). The CfA2N 
covers a volume comparable to the volume of our simulation box.
Moreover, it provides the largest published catalogue of galaxy
groups. We compare the properties of groups in the CfA2N with those of
groups extracted from a variety of mock catalogues. In addition, we 
compare the redshift space correlation functions and the pairwise
velocity dispersions of the real and artificial redshift surveys. 

Because our predicted luminosity functions are sensitive to
variations of model parameters within their plausible range, it is
important to analyse small-scale clustering in
ways that are insensitive to the exact luminosities assigned to
galaxies. To gain more insight into this question, we compare results for 
our fiducial models with results from mock surveys
of models in which galaxy luminosities are adjusted to reproduce
the CfA luminosity function exactly. (The luminosity {\it ranking} 
of the simulated galaxies is preserved during this adjustment.)  
Only the abundance of groups and the fraction of galaxies in groups 
are strongly affected by this change.  Even when the CfA luminosity 
function is reproduced exactly, some differences in small-scale clustering 
remain between $\Lambda$CDM and $\tau$CDM, and between both and the 
CfA2N. Such differences may thus provide additional constraints 
on models.

Our work is the first attempt to compare a wide-angle redshift survey 
with simulations where the physics of the formation and evolution of 
individual galaxies is treated explicitly.
Previous work has compared the CfA redshift space correlations and
group properties with $N$-body simulations either by assigning galaxies 
to dark matter particles according to some
high-peak statistical model, or by assigning them to halos based
on an assumed halo mass-to-light ratio 
(\cite{Nolthenius87}; \cite{Moore93};
\cite{Frederic95b}; \cite{Nolthenius97}; \cite{SPN97b}).
Some recent studies have attempted to simulate galaxy formation in detail
but have been forced either to treat volumes which are too small to
obtain reliable clustering statistics (e.g. \cite{Weinberg98}; 
\cite{Jenkins98}) or to use a resolution which is too poor to
follow the formation and evolution of individual galaxies
(e.g. \cite{Blanton98}; \cite{Cen98}).
In our approach, the physical processes important in  galaxy
formation are treated in a simplified way, but are included {\it ab initio}
and are followed throughout the evolution of our models. Quantities
such as the galaxy luminosity function are thus predictions of
our scheme, rather than being imposed as part of the modelling. In
Paper I, we show how differing assumptions
about star formation and feedback affect predictions for such
quantities.

In Sect. \ref{sec:3dgroup}, we analyse simulated three-dimensional (3D) groups, defined
as sets of galaxies each occupying a single dark halo, and we discuss
biases in the spatial and kinematic distributions of the galaxies
relative to those of the dark matter.  Sect. \ref{sec:mock} then describes
how we extract mock redshift surveys from our simulations, while
in Sect. \ref{sec:rsgroup} we make group catalogues from these mock 
surveys and investigate how their properties and their relation to
``real'' 3D groups are affected by the parameters which define them.
In Sect. \ref{sec:comp} we compare our simulated groups with groups extracted
in the same way from the CfA2N, and we study how this comparison is affected
by differences between the simulated and observed luminosity functions.
Finally, in Sect. \ref{sec:csi}, we calculate redshift space
correlation functions $\xi(r_p,\pi)$ and pairwise velocity dispersions
$\sigma_{12}(r_p)$ from our mock catalogues and compare with
the CfA2N. We conclude in Sect. \ref{sec:disc}.

\section{GROUPS IN REAL SPACE}\label{sec:3dgroup}

As discussed in Paper I, we identify dark matter halos in our simulations using a
friends-of-friends groupfinder which links particles closer than 20\%
of the mean interparticle separation. The most bound particle in each halo
is taken as its centre, and we calculate a radius $r_{200}$ within which
the average mass density is 200 times the critical density.
A 3D group of galaxies is then defined as a set of
three or more galaxies brighter than some chosen absolute magnitude
which lie within $r_{200}$ of a particular halo centre.

Groups in the real Universe are identified in redshift surveys. 
Below, we will compile mock surveys
from our simulations and catalogue groups identified in redshift space. 
We will then be able to address two issues: 
(1) how well the physical properties of redshift space groups correspond to those of
3D groups; (2) how well the simulation groups reproduce 
the properties of groups in the real Universe.

In this section, we describe the physical properties of the 3D groups. We
compare these with the corresponding properties of redshift space
groups in Sect. \ref{sec:linking}.

\subsection{Galaxy Luminosity Functions within Groups}\label{sec:LF}

\begin{figure}
\plotfiddle{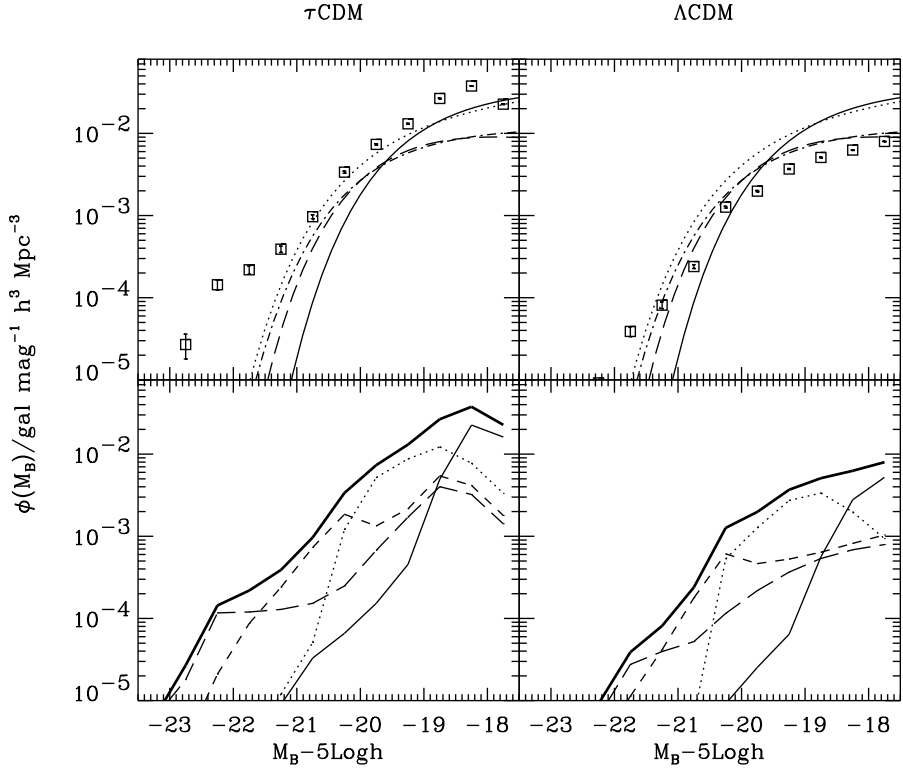}
           {0.4\vsize}              
           {90}                
           {60}                 
           {60}                 
           {210}               
           {-40}                
\caption{{\bf Upper panels}: Blue band galaxy luminosity function for our models. Error bars
are Poisson 1-$\sigma$ standard deviations. 
Solid, long-dashed, dot-dashed, and dotted lines are the CfA (\cite{MarzkeLF94}), LCRS (\cite{Lin96}),
Stromlo-APM (\cite{Loveday92}), and ESP (\cite{Zucca97}) luminosity functions, respectively.
{\bf Lower panels}: Luminosity function of galaxies within halos of different mass:
long-dashed line, $M_{200}>10^{14}h^{-1}M_\odot$; short-dashed line, 
$10^{13}h^{-1}M_\odot<M_{200}\le10^{14}h^{-1}M_\odot$; dotted line, 
$10^{12}h^{-1}M_\odot<M_{200}\le10^{13}h^{-1}M_\odot$; thin solid line, 
$M_{200}\le10^{12}h^{-1}M_\odot$. The bold solid line is the total luminosity function.}
\label{fig:lum_fun_cl}
\end{figure}

We follow galaxy formation and evolution only in dark matter halos containing at least ten 
dark matter particles. This resolution limit implies that our galaxy catalogues are
complete to a blue band absolute magnitude $M_B$ of about $-17.5 +5{\rm Log}h$. 
The luminosity functions of galaxies 
brighter than this limit are shown in Fig. \ref{fig:lum_fun_cl} for
our two fiducial models. Solid,
long-dashed, dot-dashed, and dotted lines are Schechter function fits to the
luminosity functions of the CfA Redshift Survey (Marzke, Huchra \& Geller 1994), 
LCRS (\cite{Lin96}), Stromlo-APM (\cite{Loveday92}), 
and ESO Slice Project (\cite{Zucca97}), respectively.
Table \ref{tab:lum_fun} lists the parameters of Schechter 
function fits to the fiducial models and compares them to those of the
CfA survey.  As discussed in Paper I, our
decision to normalise the models to the I-band Tully-Fisher relation
results in both models producing too many bright galaxies. 
The $\tau$CDM model also produces a faint end slope which is steeper
than the data and is a substantially worse fit to the CfA than
$\Lambda$CDM. The total blue luminosity density $\langle L_B\rangle =
\phi^*\Gamma(\alpha+2)L^*_B$ is $6.6\times 10^8 hL_\odot {\rm Mpc}^{-3}$ for the
$\tau$CDM model and $1.1\times 10^8 hL_\odot{\rm Mpc}^{-3}$ for the $\Lambda$CDM model.
Marzke et al. (1994) 
find $(2.0\pm 0.9)\times 10^8 hL_\odot{\rm Mpc}^{-3}$ for the 
CfA sample. We will see that this
disagreement between the simulated and the observed
luminosity functions affects the inferred properties of groups.

Fig. \ref{fig:lum_fun_cl} also shows luminosity functions for galaxies within halos of 
differing mass $M_{200}$ within the virial radius $r_{200}$.
Less massive halos ($M_{200}\le 10^{13} h^{-1}M_\odot$, thin solid 
and dotted lines) contribute mainly to the faint end of the luminosity function. 
Bright galaxies come from large halos ($M_{200}> 10^{13} h^{-1}M_\odot$, long-dashed 
and short-dashed lines). Note that these halos
show a slight bump in the luminosity function at $M_B-5{\rm
Log}h\sim -20.5$ and $-22$, similar to the bump in  
the luminosity function of several real clusters (see e.g. \cite{Trentham98}; \cite{Koranyi98}; 
\cite{Molinari98}) including Coma (\cite{Biviano95}). In the models, this effect
originates from the large merger cross-section of the bright and massive 
central galaxies which tend preferentially to accrete galaxies of somewhat
lower mass. 

\subsection {Cluster Profiles and Biases}\label{sec:biases}

Dynamical studies of groups and clusters have always produced mass-to-light ratios $M/L$
which are substantially smaller than that needed globally to close the Universe
(e.g. \cite{Carlberg96}; \cite{Ramella97}). 
However, it is not clear that the galaxy populations in groups are representative  
of the Universe as a whole (see Paper I). In addition, if galaxies are more strongly
clustered than the dark matter, multiplying the mean luminosity
density by the typical $M/L$ for groups and clusters can give a highly biased
estimate of $\Omega_0$. Indeed, previous work analysing mock
surveys drawn from simulations has suggested that high and low
density universes produce group catalogues which are not only almost
indistinguishable in their observational properties, but also quite similar
to the CfA groups (\cite{Nolthenius87}; \cite{Moore93}; 
Nolthenius, Klypin \& Primack 1997). 

\begin{figure}
\plotfiddle{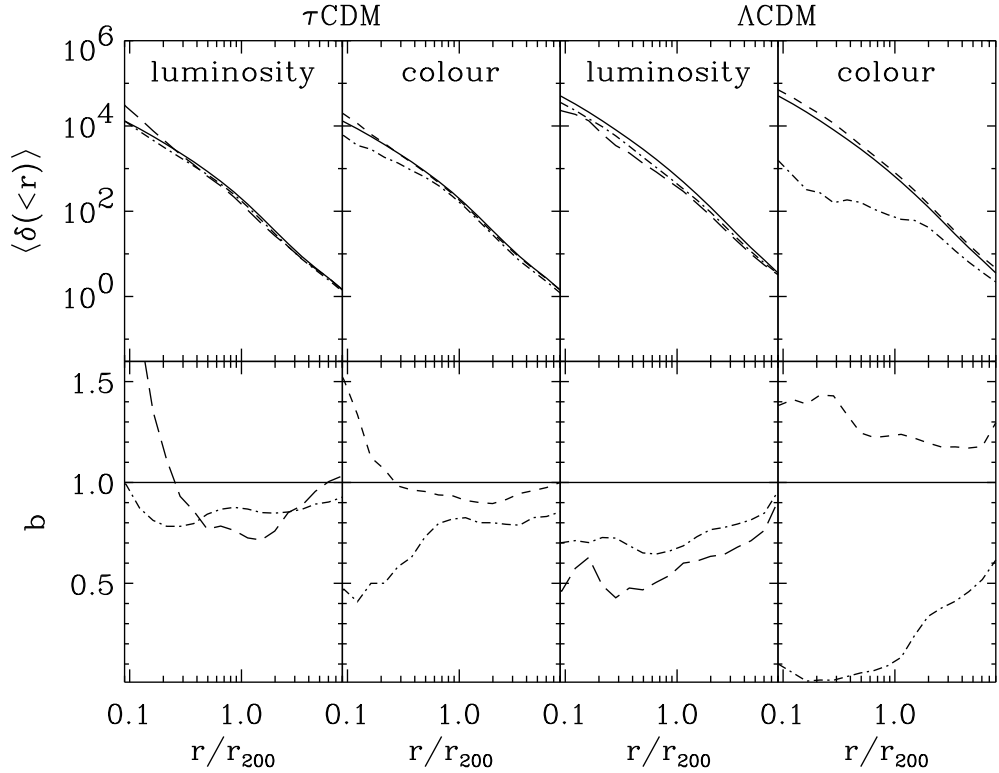}
           {0.4\vsize}              
           {90}                
           {60}                 
           {60}                 
           {210}               
           {-40}                
\caption{Average number overdensity profiles $\langle\delta(<r)\rangle$ 
and bias $b$ of galaxies relative to 
dark matter in halos with mass $M_{200}>10^{14}h^{-1}M_\odot$. 
Central galaxies of each halo are excluded in the computation of the profiles.
The two left most (right most) columns are for the $\tau$CDM ($\Lambda$CDM) model. 
{\bf First and third column}: solid, 
dot-dashed, 
and long-dashed lines are for dark matter 
and galaxies brighter than $M_B-5{\rm Log}h=-17.5$, 
and $-19.5$, respectively. 
{\bf Second and fourth
column}: solid and dot-dashed (short-dashed) lines are for dark matter and galaxies 
brighter than $M_B=-17.5+5{\rm Log}h$ and with colour $B-I\le 1.75$ ($B-I>1.75$), respectively. 
The colour cut is at the median of the colour distribution which is the same for
both cosmologies.}
\label{fig:den_bias}
\end{figure}

To investigate the distribution of galaxies relative to dark matter
in rich clusters, Figs. \ref{fig:den_bias} and \ref{fig:vel_bias} compare 
profiles of density contrast and velocity dispersion for the dark
matter and for galaxies of differing luminosity and
colour in halos with $M_{200}> 10^{14}h^{-1}M_\odot$. 
In order to compute average profiles for the whole halo
sample, we normalise the positions and the velocities of galaxies and dark matter particles
to the halo virial radius $r_{200}$ and to the circular velocity $V_{200}$ at $r_{200}$,
respectively. We then compute the average profile by superposing all
the halos in our sample and giving equal weight to each galaxy.
Note that each halo contains  a galaxy at its centre.
If we keep these central galaxies when computing the average profile, every halo 
would contribute a galaxy; we would thus obtain a galaxy number overdensity profile 
several times larger than 
the dark matter profile at radii $r<0.1 r_{200}$, only because of our arbitrary
choice of placing a galaxy at the centre of each dark matter halo. In real clusters,
this might not be the case. Therefore, 
we exclude these central galaxies when computing the profiles.
This exclusion has no effect at the radii $r\gtrsim 0.1 r_{200}$ plotted in
Figs.  \ref{fig:den_bias} and \ref{fig:vel_bias}.

The upper part of Fig. \ref{fig:den_bias} shows mean number overdensity profiles 
$\langle\delta(<r)\rangle$ for the dark matter and for galaxy samples
split by luminosity and by colour. The lower part gives the 
bias $b$, the ratio between the galaxy and the dark matter 
overdensities. In most cases the galaxies are less
concentrated than the dark matter, resulting in $b$ values less than 1
outside the cluster core. There is no strong dependence of bias on
luminosity but there is a variation with colour. We divide galaxies into a red
and a blue sample at the median of the colour distribution.
Red galaxies are more clustered than blue ones, in agreement with the
observed morphology-density relation 
(\cite{Oemler74}; \cite{Dressler80}; \cite{Postman84}; \cite{Whitmore92}; see also
\cite{Dressler97}). 
Blue galaxies are strongly anti-biased ($b<1$) in the $\Lambda$CDM
model, showing relatively little concentration towards the centres of
these rich clusters. This is a result of our assumption that galaxies
lose their gaseous halos, and so their reservoir of new gas, when they
fall into a cluster. They then redden as their interstellar medium is
used up and their star formation rate drops. The effect is much weaker
in $\tau$CDM because infall of galaxies continues at a high
rate until $z=0$ in this model. Notice that the number density
profiles for the total galaxy population have identical shapes in the 
two models but are offset in normalisation by about a factor of 3.

It is interesting to compare these results with observed number
density profiles $\Sigma(R)$ derived from 
the CNOC cluster sample (\cite{Carlberg97}) and the ESO Nearby Abell Cluster
Survey, ENACS (\cite{Biviano97}; \cite{deTheije98}). 
To obtain surface density profiles from the models, we 
assume spherical symmetry and integrate the three-dimensional profiles 
along a line of sight. At radii $\gtrsim 2r_{200}$, where
the number surface density reaches the background value, we approximate
the three-dimensional galaxy number density profile with a \cite{Navarro97} profile.
In fact, at these radii, galaxies follow the dark matter
distribution closely (Figure  \ref{fig:den_bias}). 
The absolute normalisations are difficult to establish from the observational data. We therefore 
focus on profile shapes and on the relative distribution of galaxies with different properties. 
In fact, the observational surveys find different 
concentrations between red and blue galaxies (CNOC), and between emission-line
and non-emission-line galaxies (ENACS). 

\begin{figure}
\plotfiddle{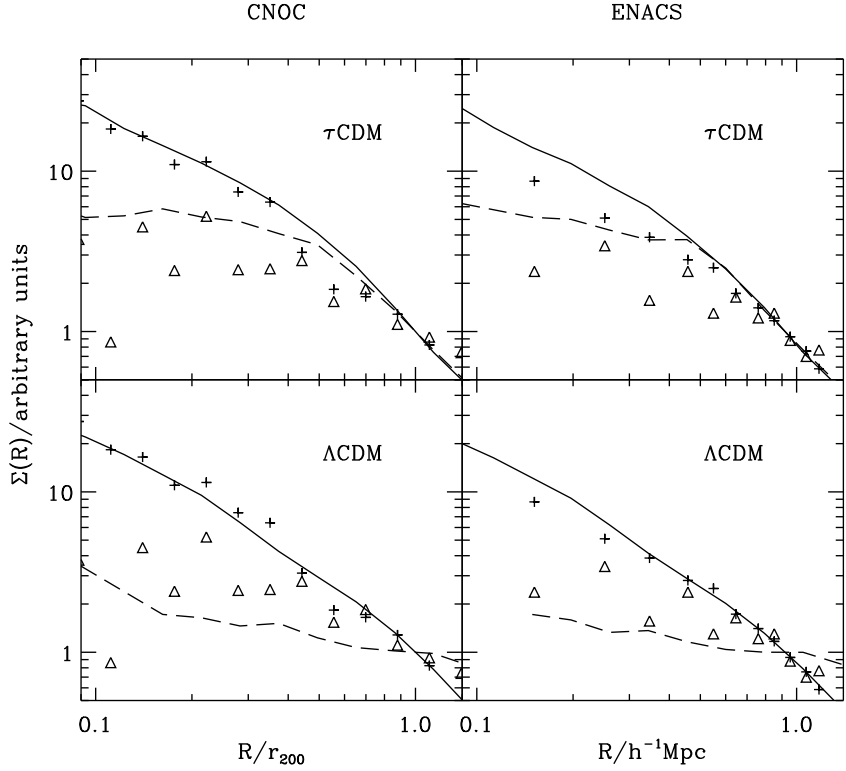}
           {0.4\vsize}              
           {90}                
           {60}                 
           {60}                 
           {210}               
           {-40}                
\caption{Surface number density profiles of galaxies within halos with mass 
$M_{200}>10^{14}h^{-1}M_\odot$ compared with observed cluster samples as explained
in the text. Solid (dashed) lines show the profiles for red (blue) galaxies
in the models. Crosses and triangles are for red and blue galaxies in the 
CNOC sample (left panels) and for non-emission-line galaxies
and emission-line galaxies for the ENACS sample (right panels). The 
fraction of galaxies in each subsample is the same for both models and
observations.}
\label{fig:surfden}
\end{figure}

\begin{figure}
\plotfiddle{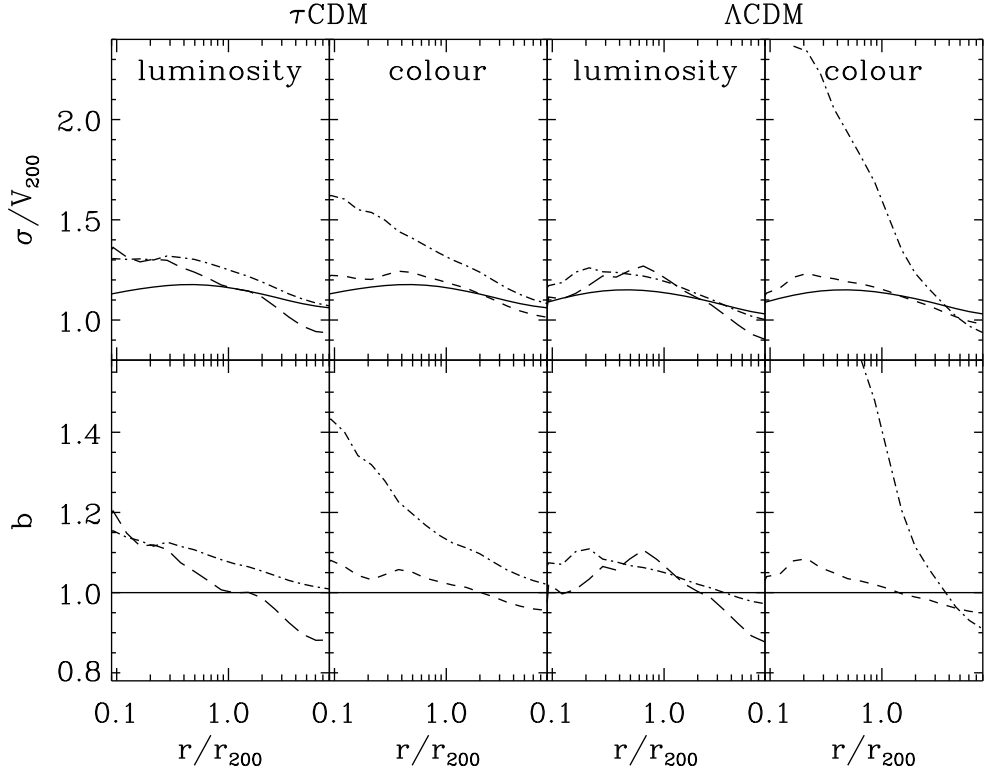}
           {0.4\vsize}              
           {90}                
           {60}                 
           {60}                 
           {210}               
           {-40}                
\caption{Velocity dispersion profiles in units of the circular velocity
$V_{200}$ and velocity bias of galaxies relative to the dark matter. Panels and
lines are as in Fig. \ref{fig:den_bias}.}
\label{fig:vel_bias}
\end{figure}

In Fig. \ref{fig:surfden}, we compare these surveys with our models. In the CNOC
sample $\sim 70\%$ of the galaxies belong to the red subsample. In the ENACS
sample the non-emission-line and emission-line subsamples 
contain $\sim 85\%$ and $\sim 15\%$ of the galaxies, respectively.
We split our simulated galaxy samples into red and blue subsamples in such a way as to 
reproduce these fractions.
We normalize both the observed and the model profiles at $\Sigma(r_{200})$. For the ENACS
sample, we assume $r_{200}=0.92 h^{-1}$ Mpc, the mean value of $r_{200}$ for the
clusters in both our $\tau$CDM and $\Lambda$CDM models. 

These surveys give mean profile shapes very similar to those we find here. 
In addition, the differences between the observed subsamples 
are intermediate in strength between those predicted by our two models.
Note however that the results of Fig. \ref{fig:surfden} are only
indicative, because the relative distribution of galaxy subsamples is
sensitive to the selection criteria. For example, when the spectra are used to split 
the ENACS sample into early-type non-emission-line and late-type emission-line galaxies, 
the difference widens (\cite{deTheije98}). 

Finally, Fig. \ref{fig:vel_bias} shows, for our models, the velocity dispersion (in units of the
circular velocity $V_{200}$) for galaxies within spheres of radius $r$
(in units of $r_{200}$). These dispersions depend very little on the
luminosity of the galaxies and are very similar to the dark matter dispersions.
There is a noticeable colour effect, however, which is quite large 
in the $\Lambda$CDM model. The velocity dispersions of the red
galaxies trace those of the dark matter, whereas blue galaxies have
dispersions a factor $\approx 1.5-2.0$ larger. A similar effect is
seen in real clusters (\cite{Moss77}; \cite{Mohr96};
\cite{Carlberg97}; \cite{Biviano97}, 1998); it reflects the fact that the few observed blue
galaxies have just fallen into the cluster and so are more weakly
bound than the red galaxies.

The weak velocity bias we find for most galaxy samples is consistent with
the similarity between the moments of the pairwise velocity
distribution for the galaxies and for dark matter (see
Paper I).

\subsection {Dynamical Properties of 3D Groups}\label{sec:3d-dyn}

Table \ref{tab:gr_3D} lists the quartiles of 
the distributions of a variety of 3D group properties. Groups are here
defined as halos which contain three or more galaxies brighter than 
$M_B=-17.5+5{\rm Log}h$ within $r_{200}$.

The harmonic radius $R_h$, the one-dimensional 
velocity dispersion, $\sigma$, and the total luminosity, $L_B$ of
galaxy groups are computed directly using the positions, velocities and luminosities
of the galaxies. Assuming virial equilibrium, we can combine these
to derive a virial mass estimate, $M_{\rm vir}$, and 
so an estimated virial mass-to-light ratio, $M_{\rm vir}/L_B$. 
(We review our definition of these quantities in the
Appendix. Note that $L_B$ includes a correction for
galaxies fainter than our absolute magnitude limit. On average, $\sim 50\%$ and $\sim 30\%$
of the total halo luminosity comes from this correction factor in 
the $\tau$CDM and $\Lambda$CDM model respectively.)  In the 
simulations the virial mass estimate can be compared
with the {\it true} mass of the group $M_{200}$.

The first thing to notice from Table \ref{tab:gr_3D} is that 
groups are very similar in the two models. They have slightly larger sizes,
lower luminosities and larger velocity dispersions in $\Lambda$CDM;
these differences combine to give typical $(M_{\rm vir}/L_B)$
values which are two times {\it larger} in the low density model.
This surprising result can be attributed primarily to the lower
luminosity density in the $\Lambda$CDM model. If we multiply the median
$(M_{\rm vir}/L_B)$ of the groups in each simulation by the mean luminosity 
density, we can estimate $\Omega_0$ by dividing the result by the appropriate
critical density. These estimates $\Omega_0^{\rm est}$ are given in the last line of Table
\ref{tab:gr_3D} and in both cases are nearly a factor of 2 smaller
than the actual value of $\Omega_0$. Notice that $M_{\rm vir}$ actually
overestimates the true group mass systematically.
If we use the true median $M_{200}/L_B$ of groups 
to compute $\Omega_0^{\rm est}$ rather than our median virial estimate, we
get 0.40 and 0.13 for $\tau$CDM and $\Lambda$CDM
respectively. Clearly, the standard assumption that groups have the
same mass-to-light ratio value as the universe as a whole is untrue in either
model. The bias in mass-to-light ratio is actually a function of
group mass as can be seen in Fig.~15 of Paper I.

\section{MOCK CATALOGUES}\label{sec:mock}

The simulation boxes have  volume $\sim 6\times 10^5 h^{-3}$ Mpc$^3$ and 
$\sim 2.8\times 10^6 h^{-3}$ Mpc$^3$ for the $\tau$CDM and the $\Lambda$CDM models, respectively.
Both simulations are normalised to match the present day cluster abundance. 
As a result both simulation
boxes contain several clusters of mass $\sim 10^{15} h^{-1} M_\odot$. The CfA2N 
has volume $\sim 7\times 10^5 h^{-3}$ Mpc$^3$ within
$cz=12000$ km s$^{-1}$ and contains four Abell clusters of richness
$R=2$ (including the Coma cluster) with masses $\sim 10^{15} h^{-1} M_\odot$.
The CfA2N covers the right ascension range $[8^h,17^h]$ and
the declination range $[8.5^{\rm o},44.5^{\rm o}]$, thereby avoiding
regions of strong obscuration. It is clearly well suited for comparison with the
simulations.

In Sect. \ref{sec:LF} we showed that the simulated luminosity functions
agree rather poorly with the observations.                                    
The discrepancy with the                           
CfA luminosity function (\cite{MarzkeLF94}) is particularly large because
the CfA survey is unusually dense: the normalisation
$\phi^*$ is two times larger and the characteristic magnitude $M_B^*$
is $\gtrsim 0.5$ mag fainter than the average values derived from other surveys.
Furthermore, the CfA2N is significantly denser than the CfA survey as
a whole, presumably because the ``Great Wall'' dominates this region.

In order to understand how the luminosity functions of our models affect the properties of
the groups and the redshift space correlation functions we study
below, we have constructed mock catalogues from our simulations in two different ways:
\begin {enumerate}
\item  We keep the luminosities of the galaxies
 predicted by the semi-analytic recipes; we refer to catalogues made
 using these luminosities      
 as semi-analytic luminosity function (SALF) catalogues; 
\item  We alter the luminosities of the galaxies in the simulations  
 in order to obtain an exact match to the CfA luminosity
 function. This is done as follows. For each model we select
 a set of luminosities from the CfA luminosity function constrained
 so that their number equals the number of simulation galaxies
 with original luminosity $M_B\le -17.5+5{\rm Log}h$ and
 their sum equals the simulated volume times the CfA luminosity
 density. The new luminosities then replace the original ones in such
 a way that the luminosity ranking of the galaxies is unaltered.
 For each galaxy we compute the shift from the original to the new luminosity.
 Ninety per cent of the luminosity
 shifts lie in the ranges $[0.5,1.0]$ and $[-0.9,0.0]$ magnitudes
 in the $\tau$CDM and $\Lambda$CDM models respectively;
 the median shifts are 0.8 and $-0.5$ magnitudes respectively.
 We refer to catalogues constructed using these new luminosities as 
 CfA luminosity function (CfALF) catalogues. 
 \end {enumerate}

The large scale structure of CfA2N is dominated by the
Great Wall and the Coma cluster.
The center of Coma has celestial coordinates $\alpha_{1950}= 12^h 57^m$, 
$\delta_{1950}= 28.32^{\rm o}$
(see e.g. \cite{Gurzadyan98}) and its distance from the Milky Way is $\sim 70 h^{-1}$ Mpc.
To compile mock catalogues we place a hypothetical observer at a distance 
$d\in [68,72]h^{-1}$ Mpc from the most massive cluster within the simulation
box. The location of the observer is chosen to  coincide with the position of a 
galaxy with properties
similar to the Milky Way, i.e. a  blue band luminosity
$L_B\sim 1.9\times 10^{10}L_\odot$ and mass-to-light ratio $\sim 5 M_\odot/L_\odot$ 
(e.g. \cite{Gilmore90}). If these values are valid for a fiducial
Hubble parameter $H_0=65$ km s$^{-1}$ Mpc$^{-1}$, we have $M_B=-19.3 +5{\rm Log}h$.
We thus look for a galaxy with magnitude $M_B-5{\rm Log}h\in[-18.9,-19.7]$, stellar mass
$M\in [4,8] \times 10^{10} h^{-1}M_\odot$, 
and star formation rate SFR$\in[0.1,10]M_\odot$ yr$^{-1}$, 
typical of a normal spiral galaxy (\cite{Kennicutt98}). 
We also require that this galaxy belongs to a dark matter halo similar
to the Local Group halo, with total mass $M\in[0.1,1.0]\times 10^{13} h^{-1} M_\odot$ and
containing no more than 2 galaxies brighter than $M_B-5{\rm Log}h=-17.5$, the magnitude 
of M33. We find 32 and 15 galaxies satisfying these criteria in 
the $\tau$CDM and $\Lambda$CDM model respectively. 
  
Once we have found the observer's galaxy, we rotate the reference frame so that
the massive cluster has the coordinates of Coma. We then compile a catalogue of galaxies
with the same right ascension and declination range as the CfA2N.
To each galaxy we assign the radial velocity $cz=({\bf v}-{\bf v}_{hg})\cdot {\bf r}/r
+{\bf r}$, where ${\bf v}$ is the galaxy peculiar velocity,
${\bf v}_{hg}$ is the observer's peculiar velocity, ${\bf r}$
is the relative position between the galaxy and the observer and 
is in units of km s$^{-1}$.
The blue absolute magnitude $M_B$ 
yields the apparent magnitude $m_B=M_B+25+5{\rm Log}(r/h^{-1}{\rm Mpc})$. 
We include all galaxies
with $cz\in[500,15000]$ km s$^{-1}$ and $m_B\in[10,m_{\rm lim}]$. 
Redshift and magnitude lower cutoffs avoid including faint objects close to the home galaxy.

We choose the magnitude limit $m_{\rm lim}$ of our mock redshift surveys as follows. 
The CfA2N catalogue has a Zwicky magnitude limit (roughly a $B$-band magnitude limit) 
$m_{\rm lim}=15.5$.  Because the 
semi-analytic  and the CfA survey luminosity densities differ by more than a factor of two, 
we fix $m_{\rm lim}$ in the SALF catalogues  by requiring that the
number of galaxies in the mock survey be  $\sim 6000$, 
the number of galaxies within CfA2N. We obtain  $m_{\rm lim} \sim 15$ for the $\tau$CDM model  
and $m_{\rm lim}\sim 16$ for $\Lambda$CDM. 
Errors in magnitude estimates of the CfA2N catalogue 
and the scatter in the correlation between blue and Zwicky magnitudes
make a shift in  $m_{\rm lim}$ by $\approx 0.5$ mag not at all unreasonable (see
\cite{MarzkeLF94} and references therein). 
The CfALF catalogues have the CfA luminosity function and luminosity density by 
definition, so, in this case, we set $m_{\rm lim}=15.5$.

Note that our simulation box is $85$ ($141$) $h^{-1}$ Mpc on a side 
for the $\tau$CDM ($\Lambda$CDM) model.
Thus, in order to have a survey with a depth of  $15,000$ km s$^{-1}$,
one periodic replication of the simulation box is required.
For  $m_{\rm lim}=15.5$, galaxies more distant than
85 (141) $h^{-1}$ Mpc are always brighter than $M_B-5{\rm Log} h=-19.15$ ($-20.25$)
for $\tau$CDM ($\Lambda$CDM); only the brightest
galaxies enter the mock catalogue from the replicated regions. Since we
concentrate in this paper on small scale clustering, we do not expect
our results to be seriously affected by this limitation.

\begin{figure}
\plotfiddle{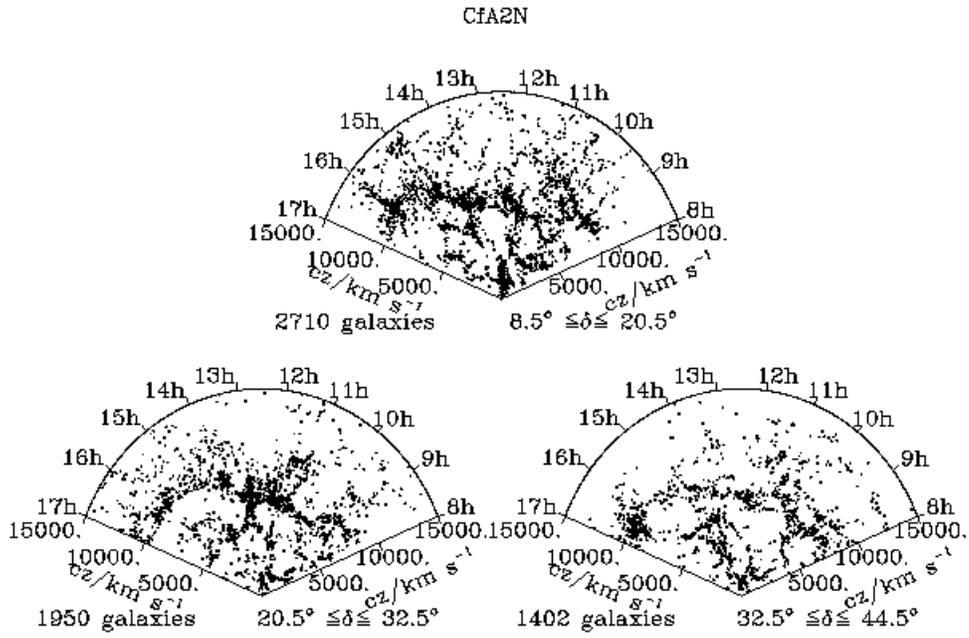}
           {0.4\vsize}              
           {90}                
           {60}                 
           {60}                 
           {210}               
           {-40}                
\caption{Galaxy distribution in the CfA2N catalogue projected onto three declination
intervals.}
\label{fig:slice_cfa2n}
\end{figure}

\begin{figure}
\plotfiddle{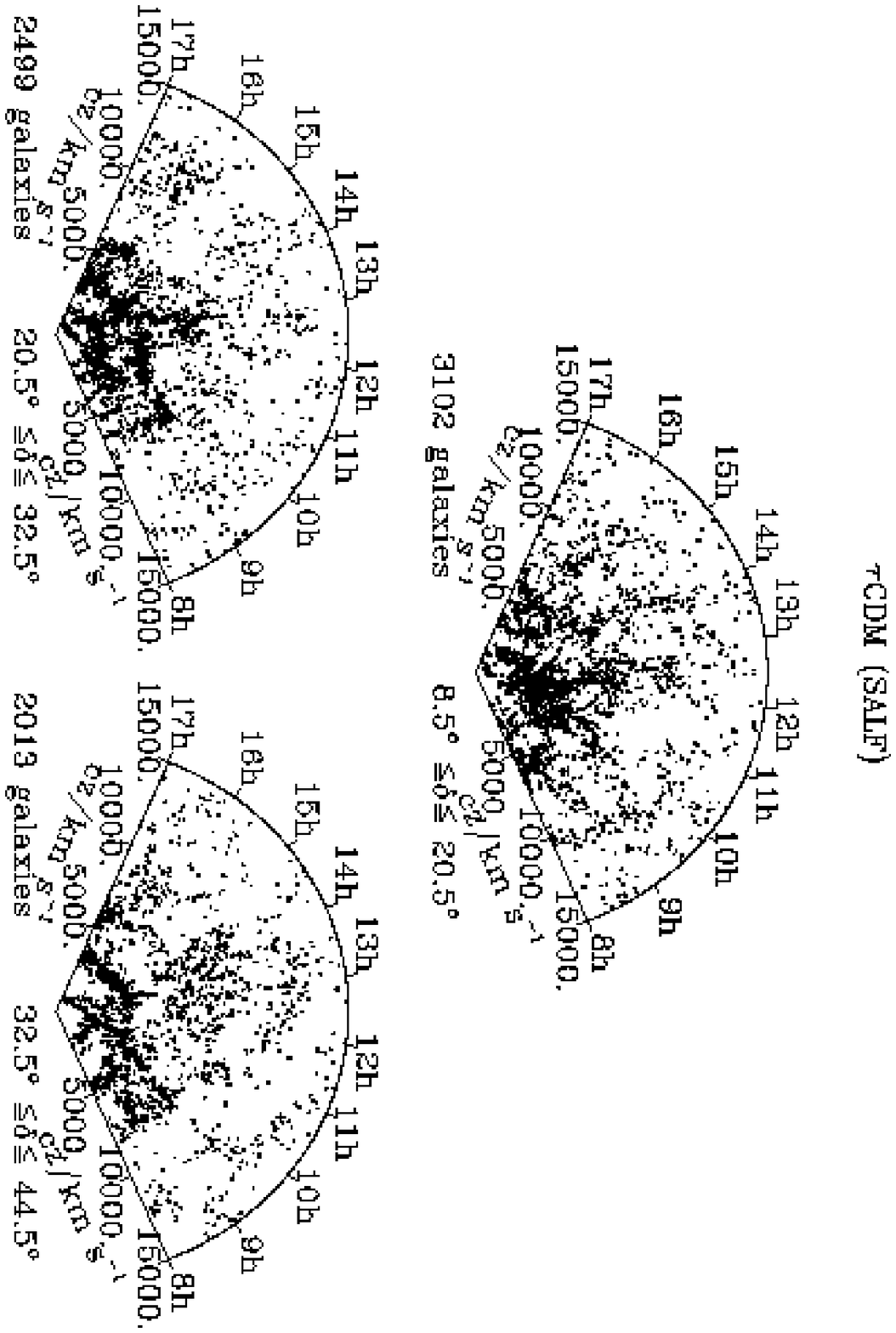}
           {0.4\vsize}              
           {90}                
           {60}                 
           {60}                 
           {210}               
           {-40}                
\caption{Same as Fig. \ref{fig:slice_cfa2n} for a SALF catalogue extracted 
from the $\tau$CDM simulation box.}
\label{fig:slice_tcdm}
\end{figure}

\begin{figure}
\plotfiddle{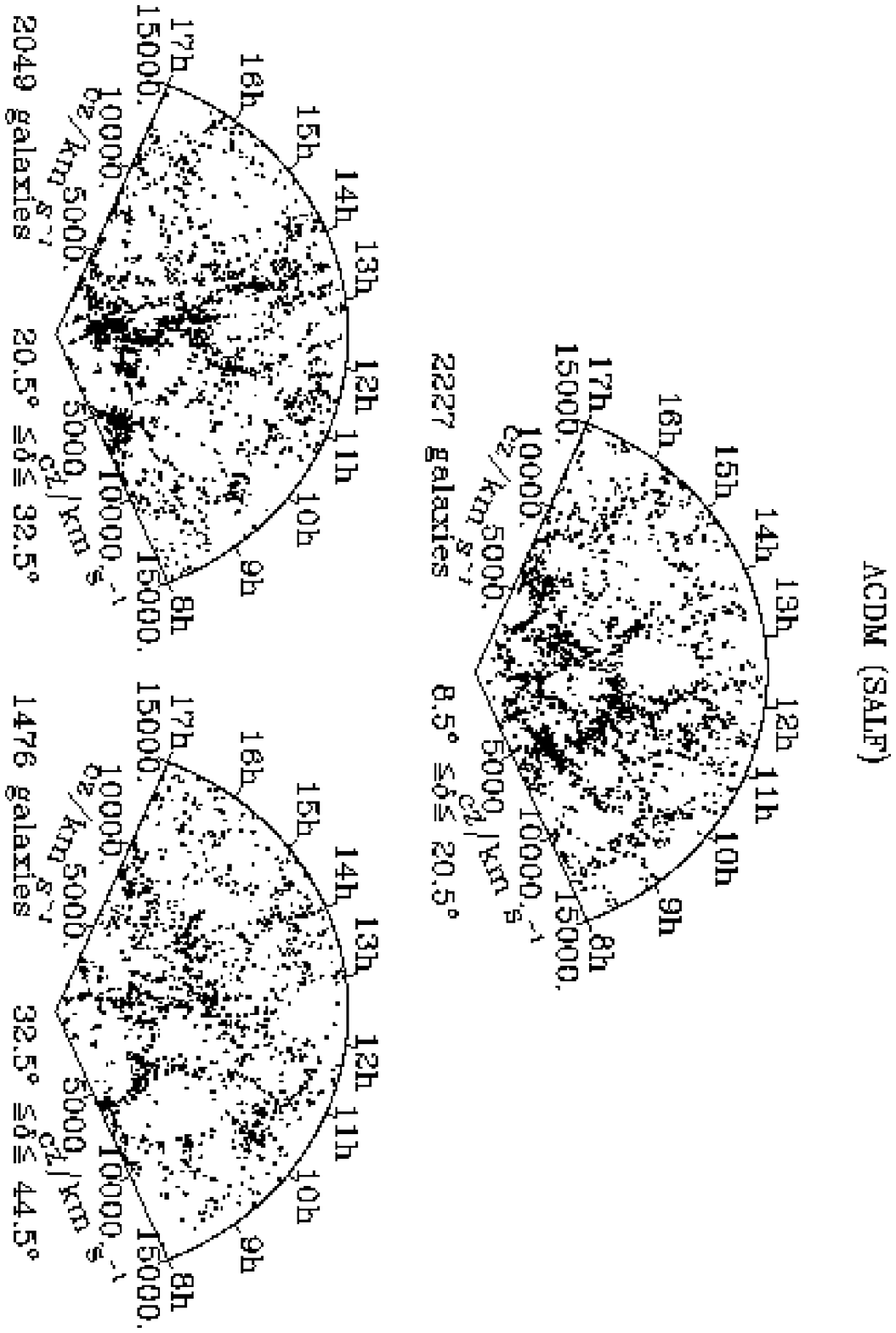}
           {0.4\vsize}              
           {90}                
           {60}                 
           {60}                 
           {210}               
           {-40}                
\caption{Same as Fig. \ref{fig:slice_cfa2n} for a SALF catalogue extracted from 
the $\Lambda$CDM simulation box.}
\label{fig:slice_lcdm}
\end{figure}

\begin{figure}
\plotfiddle{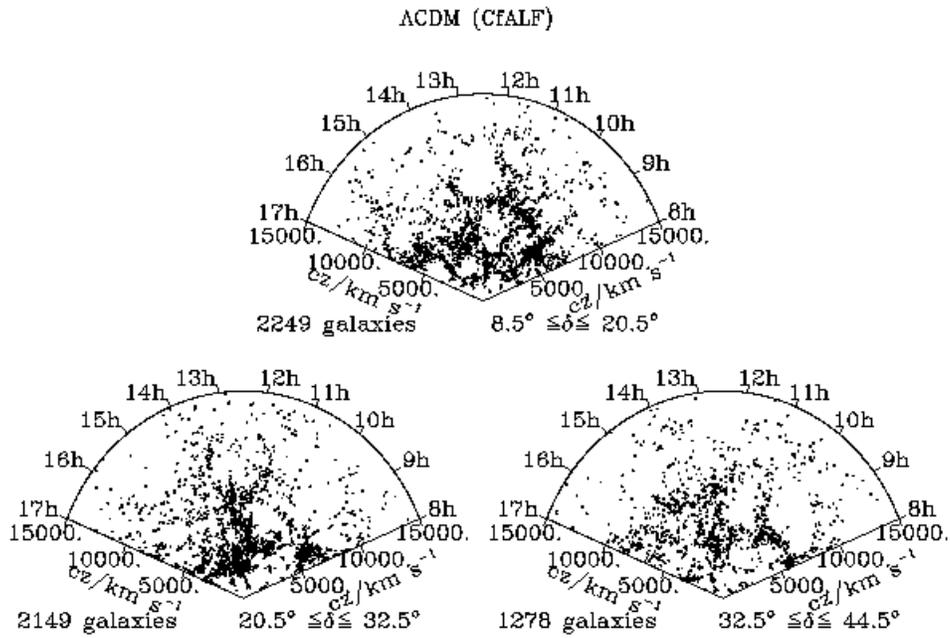}
           {0.4\vsize}              
           {90}                
           {60}                 
           {60}                 
           {210}               
           {-40}                
\caption{The CfALF catalogue corresponding to the SALF catalogue shown in Fig. \ref{fig:slice_lcdm} 
extracted from the $\Lambda$CDM simulation box.}
\label{fig:slice_lcdm.cfa}
\end{figure}

Figs. \ref{fig:slice_cfa2n}, \ref{fig:slice_tcdm} and \ref{fig:slice_lcdm} 
show the CfA2N catalogue and two typical SALF catalogues from the $\tau$CDM and the 
$\Lambda$CDM simulations respectively. The $\tau$CDM mock catalogue does not look very much
like the real Universe. There is too much structure within 5000 km s$^{-1}$ 
and the model fails to produce 
coherent sheets or filaments. These features are common to all our $\tau$CDM
mock catalogues. The $\Lambda$CDM catalogues are in better qualitative
agreement with the data. There are large voids and filaments extending across the
full survey volume. Nevertheless, the structures are not as striking
or as sharply defined as in the CfA2N.
In addition, there is no structure comparable to the ``Great Wall''. We have searched the
simulation boxes for  coherent sheets of galaxies and have failed to find anything as large
as the observed wall (see also \cite{Schma99}). It is unclear if this is a result of the
relatively small volume of our simulations (particularly $\tau$CDM) or
reflects a significant problem for the cosmological models we have
studied. CfALF mock redshift surveys are
closer in appearance to the CfA2N because the change in
luminosity function forces a distribution of galaxies in redshift
closer to that observed (Figure \ref{fig:slice_lcdm.cfa}). Despite this the qualitative differences
in large scale structure between the three cases remain.

In the following sections, we analyse the CfA2N and mock catalogues 
extracted from the simulations using identical techniques.  
We have compiled an ensemble of ten mock 
catalogues for each simulation in order to assess the robustness of our
results. Note, however, that since all ten are constructed from the
same parent simulation, the scatter between statistics estimated from
them will underestimate the true sampling variance. 

\section{GROUPS IN REDSHIFT SPACE}\label{sec:rsgroup}

A major problem for the study of groups in the real Universe is in
finding an objective algorithm to define them, given 
that only partial knowledge of the phase space coordinates is
available in real galaxy catalogues.
This algorithm should select groups which correspond as closely
as possible to real 3D groups, at least in a statistical sense.
In the eighties, two main algorithms were proposed:
the hierarchical method (e.g. \cite{Materne78}; 
\cite{Tully80}, 1987) 
 and the friends-of-friends algorithm
(\cite{Huchra82}; \cite{Nolthenius87}).

We identify groups in our mock catalogues with the friends-of-friends algorithm
described by Ramella et al. (1997).
The robustness of methods of this type
has been studied by Nolthenius \& White (1987), 
Moore et al. (1993), 
and \cite{Frederic95a}. All of these studies used dark matter--only
$N$-body simulations to study the relation between groups selected
from mock observational catalogues and genuine virialized systems. Their
results were very similar to those we find below -- with careful
parameter choices it is possible to arrange a fair correspondence
between the two kinds of system and to ensure that the statistical properties
of groups estimated from the ``observational'' catalogues are quite similar
to those of the ``real'' systems.

Sect. \ref{sec:method} reviews the friends-of-friends algorithm. In Sect. 
\ref{sec:linking}, we investigate the dependence of the statistical
properties of groups on the linking parameters. 
In Sect. \ref{sec:comp} we compare our models with the CfA2N groups. We 
find  that variations of the  galaxy luminosity
function {\em do} affect the absolute abundances
of groups, but have very little effect on their median properties. 

\subsection{Method}\label{sec:method}

The friends-of-friends algorithm is an approximate method for
identifying systems which lie above a chosen number overdensity threshold. 
We know galaxy positions in redshift space rather than 
in real space. We therefore need to use two distinct linking lengths, $V_0$ and $D_0$, 
for the radial velocity coordinate and the coordinates projected onto the sky, respectively.
When we know the galaxy luminosity function $\phi(M)$, $D_0$ defines the number
overdensity threshold
\begin{equation}
{\delta n\over n} = {3\over 4\pi D_0^3} \left[\int_{-\infty}^{M_{\rm lim}} \phi(M)dM\right]^{-1}-1
\end{equation}
where $M_{\rm lim} = m_{\rm lim} -25-5{\rm Log}(cz_f/H_0)$ is the faintest observable 
absolute magnitude at the fiducial redshift $cz_f=1000$ km s$^{-1}$ within a redshift
survey with apparent limiting magnitude $m_{\rm lim}$. Having chosen the threshold 
$\delta n/n$ and the linking length, $V_0$, we link each pair of galaxies which satisfies
\begin{equation}
{cz_i+cz_j\over H_0}\sin\left(\theta_{ij}\over 2\right) \le D_0 R_{ij}
\end{equation}
\begin{equation}
\vert cz_i-cz_j\vert\le V_0 R_{ij}
\end{equation}
where $cz_i$ and $cz_j$ are the galaxy radial velocities, $\theta_{ij}$ is their angular
separation and 
\begin{equation}
R_{ij}=\left[\int_{-\infty}^{M_{\rm lim}} \phi(M)dM\over \int_{-\infty}^{M_{ij}} \phi(M)dM\right]^{1/3
}
\label{eq:scaling}
\end{equation} 
where $M_{ij} = m_{\rm lim} -25-5{\rm Log}[(cz_i+cz_j)/2H_0]$. 
Note that the scaling law in eq. (\ref{eq:scaling}) has been questioned by
many authors. Specifically, replacing the power $1/3$ with $1/2$ (see the 
argument in \cite{Nolthenius87}; \cite{Magtesyan88}; \cite{Gourgoulhon92}) 
drastically reduces the correlation between 
redshift and velocity dispersion observed in the Huchra \& Geller (1982) 
group catalogue. However, part of this correlation is related to a selection
effect rather than to the grouping algorithm, because 
groups with low velocity dispersion usually have few bright galaxies and so
can only be seen at low redshift.
Here, we use eq. (\ref{eq:scaling}) for consistency with the Ramella et al. (1997) 
catalogue.


One of the goals of identifying groups in the real Universe
is to estimate their number density as a function of properties such as
luminosity or velocity dispersion. 
To compute the correct abundance of groups, we weight each 
according to its distance (\cite{Moore93}). We consider groups with
$N\ge 3$ members. We can thus identify a group only when its third-ranked galaxy
has absolute magnitude $M_j\le m_{\rm lim}-25-5{\rm Log}(\langle cz\rangle/H_0)$, where
$\langle cz\rangle$ is the mean redshift of the group. $M_j$ determines
the radius $cz_j$ of the sphere within which we could have identified this group.
This group contributes with weight $1/\Psi_j$ to the total abundance of groups, where
\begin{equation}
\Psi_j={\Omega\over 3}\left(cz_j\over H_0\right)^3 \left[ 1-{3z_j\over 2}
\left(1+{\Omega_0\over 2}\right)\right]
\label{eq:gr_weight}
\end{equation}
is the proper volume sampled by the group, to first order in $z_j$, $\Omega$ is the solid angle
of the survey and $\Omega_0$ is the (unknown) cosmological density parameter. We consider
groups with $\langle cz\rangle\le 12000$ km s$^{-1}$; therefore the first order
correction in the volume  is 9\% at the most, when $\Omega_0=1$.

Note that the simulation galaxy catalogues are complete
to $M_B=-17.5+5{\rm Log}h$. This magnitude sets a minimum redshift $cz_{\rm min}$:
groups closer than $cz_{\rm min}$
could contain galaxies fainter than $M_B$ in a real magnitude limited
survey. Therefore, we consider only redshift space groups with $\langle cz\rangle \ge cz_{\rm min}$
both in the mock catalogues and in the CfA2N catalogue.

\subsection{Linking Parameters, Interlopers, and Physical Properties}\label{sec:linking}

From the CfA2N catalogue, Ramella et al (1997) 
compiled a fiducial group catalogue
with linking parameters $\delta n/n=80$, and $V_0=350$ km s$^{-1}$.
Here, we show that in the simulations these linking parameters give groups in redshift
space with similar velocity dispersions and luminosities to the 3D
groups (Sect 2.3) but with substantially larger sizes.
These linking parameters will then be adopted for the remainder of our analysis.

The accidental inclusion of interlopers is one of the major problems 
in identifying groups in redshift space.
Even with full knowledge of the galaxy locations in the six-dimensional phase space,
the concept of interloper is ill-defined because groups are not isolated. 
Here, we  define interlopers as follows. 
Each 3D group contains only galaxies lying within the same dark
halo. Each member of a redshift space group is associated with
some dark halo, but such a group may contain galaxies belonging
to different dark halos. We define the dark halo
containing the greatest number of group members as the group halo.       
All members belonging to another halo are then interlopers,
and we define the interloper fraction $f_{\rm int}$ as the ratio 
between the number of interlopers and the total number of true members 
plus interlopers. If all members belong
to different dark matter halos, the group is spurious and we set 
$f_{\rm int}=1$, because the group halo is undetermined.

\begin{figure}
\plotfiddle{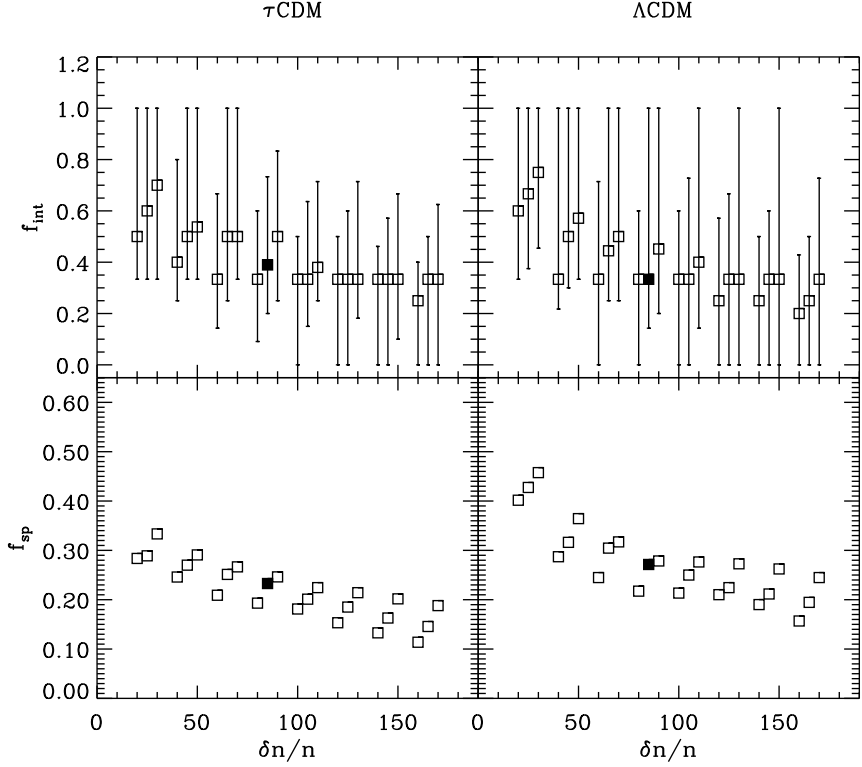}
           {0.4\vsize}              
           {90}                
           {60}                 
           {60}                 
           {210}               
           {-40}                
\caption{{\bf Upper panels}: Median of the fraction of interlopers $f_{\rm int}$ 
within an individual group in group catalogues extracted from the two SALF catalogues 
shown in Figs. \ref{fig:slice_tcdm} and \ref{fig:slice_lcdm} with different 
linking parameters $\delta n/n$ and $V_0$. Squares are the weighted medians
of the distributions, error bars show the lower and upper quartiles. At fixed
$\delta n/n$, quantities for $V_0=150$, 350, and 550 km s$^{-1}$ are
shown; quantities for $V_0>150$ km s$^{-1}$ are shifted to the right for clarity. 
Solid square refers to the fiducial catalogue. 
{\bf Lower panels}: Fraction $f_{\rm sp}$ of spurious groups.}
\label{fig:link_combmod}
\end{figure}

Fig. \ref{fig:link_combmod} 
shows the dependence of $f_{\rm int}$  on the values
of the linking parameters $V_0$ and $\delta n/n$ for two SALF 
redshift surveys.
All the CfALF and  SALF  mock catalogues we compiled 
yield similar results.  Squares show the median of
the distribution of $f_{\rm int}$; error bars show the first and
third quartiles. The solid square refers to the fiducial catalogue 
$\delta n/n=80$, $V_0=350$ km s$^{-1}$. 
It is apparent that $f_{\rm int}$ reaches a minimum when $\delta n/n\gtrsim 80-100$ and $V_0$
is small. Note that this minimum value is quite large; more than a third
of the assigned members of a typical group are interlopers. Fig. 
\ref{fig:link_combmod} also shows that a significant fraction of
groups are spurious; for the preferred values of $\delta n/n$ 
this fraction is 20 to 30\%. We find that $\sim 40\%$ of the
triplets are spurious, whereas only $\sim 20\%$ of the groups with four or more members are
spurious. This result agrees with the suggestion of \cite{Ramella89} that
the physical association of triplets should be considered uncertain.

\begin{figure}
\plotfiddle{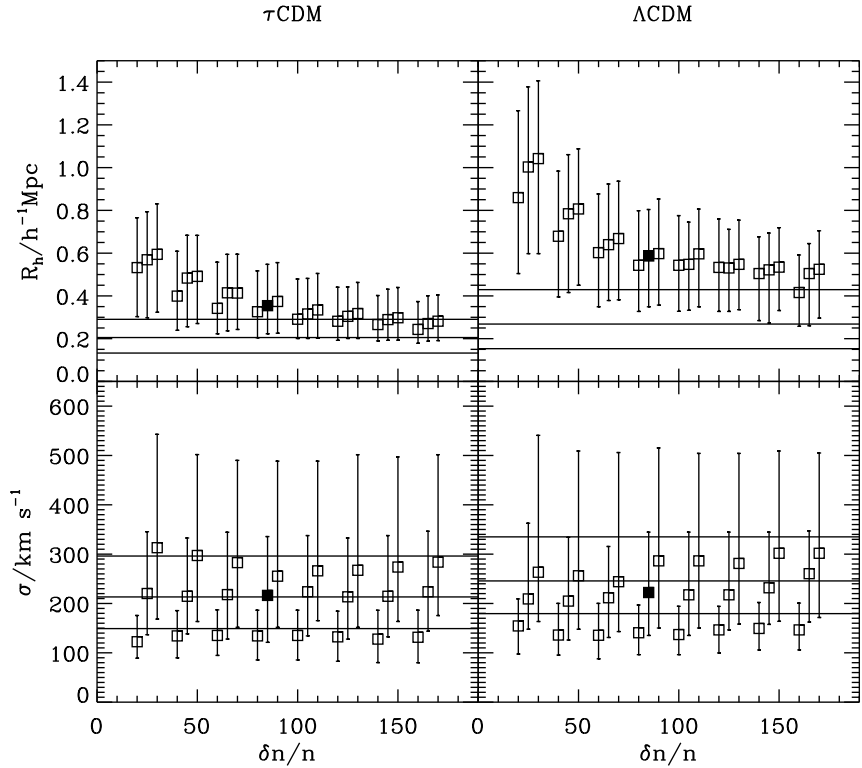}
           {0.4\vsize}              
           {90}                
           {60}                 
           {60}                 
           {210}               
           {-40}                
\caption{Dependence of the group median properties on the linking parameters
$\delta n/n$ and $V_0$ in the SALF catalogues of Figs. \ref{fig:slice_tcdm} and \ref{fig:slice_lcdm}. 
Values of $V_0$ are as in Fig. \ref{fig:link_combmod}.
Error bars show the upper and lower quartiles. Solid lines show the quartiles 
of the 3D group catalogue.}
\label{fig:gr_ratio_combmod}
\end{figure}

Fig. \ref{fig:gr_ratio_combmod} shows how the weighted quartiles
of the harmonic radius $R_h$ and of the velocity dispersion $\sigma$
vary with the linking parameters which define our catalogues. By
weighted quartiles we mean the 25, 50 and 75\% points
of the cumulative distribution when each group is
assigned weight $1/\Psi_j$ (see equation \ref{eq:gr_weight}); these
should correspond approximately to the quartiles of a volume
limited group sample. For comparison we also plot the quartiles
of the corresponding distributions for 3D groups as listed in
Table 2. These figures show that $R_h$ depends only weakly
on $V_0$ and is a decreasing function of $\delta n/n$; for
$\delta n/n\gtrsim 80$ it depends little on either linking parameter.
On the other hand, $\sigma$ is a strong function of $V_0$ for
all $\delta n/n$. This systematic behaviour has been noted in all
previous investigations of grouping algorithms; see
Trasarti-Battistoni (1998) for a recent discussion. While our
fiducial choice, $V_0=350$ km s$^{-1}$, leads to a $\sigma$
distribution which is similar to that of the true 3D groups,
taking $\delta n/n = 80$ results in $R_h$ distributions which
are biased high by about a factor of two. 

This bias in group size is related to the rather high interloper 
fraction noted above. If interlopers are excluded when calculating 
group properties, then the remaining galaxies are, by definition,
all members of the same halo and so give a measure of group size which
is statistically similar to that found using full 3D information.
We demonstrate this in Tables \ref{tab:gr_RS_tcdm} and
\ref{tab:gr_RS_lcdm},  which compare the weighted quartiles of the 
distributions in mock redshift surveys including and excluding
interlopers with those of the 3D groups repeated from Table \ref{tab:gr_3D}. 
Excluding interlopers has little effect on the
velocity dispersions of the groups but brings their estimated
sizes and mass-to-light ratios into much better agreement with those
of the 3D groups. Unfortunately, of course, it is not possible to
exclude interlopers from real redshift survey group catalogues in
this fashion.

It is curious that in our models this interloper-induced
overestimation of $R_h$ in redshift survey groups combines with 
the overestimation of true group mass by $M_{\rm vir}$ (see section
\ref{sec:3d-dyn}) to almost exactly cancel out the factor of 2.5 difference
between the true mass-to-light ratio of groups and the mean
mass-to-light ratio of the Universe. As a result, as may be seen in
Tables \ref{tab:gr_RS_tcdm} and \ref{tab:gr_RS_lcdm}, estimates
of $\Omega_0$ obtained by multiplying the (incorrect) mean
estimated $M_{\rm vir}/L_B$ by the average luminosity density are fortuitously
quite close to the true value. It is difficult to judge whether
this interloper-bias conspiracy will also work in the real
Universe. Some of the tests we carry out below
suggest that there is no reason to expect that these two
biases should always be of the same order.

\section{COMPARISON WITH THE CfA2N GROUPS}\label{sec:comp}

In this section, we compare our groups with the CfA2N catalogue. The galaxy luminosity function
is a fundamental ingredient in constructing mock catalogues and 
in the friends-of-friends algorithm, so we
might expect properties of groups in our simulated redshift surveys to
depend strongly on the adopted luminosity function.
We show below that this is only partly the case. We investigate how the luminosity
function affects group abundances (Sect. \ref{sec:real_gr}) and the
dependence of group properties on the linking parameters (Sect.
\ref{sec:cfa_link}). 

\subsection{Linking Parameters}\label{sec:cfa_link}

In this section we study how the median group velocity dispersion
$\sigma$ depends on the linking parameters used to define groups, and we
compare with the behaviour seen in the CfA2N survey.
The harmonic radius $R_h$ and the total group luminosity $L_{\rm tot}$
are also directly  measurable, but, as we have shown, the
estimated value of $R_h$ is biased high by interloper contamination, while
$L_{\rm tot}$ is  insensitive to the linking parameters.
On the other hand,  the velocity dispersion $\sigma$ is not
significantly biased by interloper contamination and is sensitive to the linking
parameter $V_0$. Moreover, the median $\sigma$ is a direct measure of the
dynamics of galaxies on small scales, which in turn depends on the
underlying mass distribution.

\begin{figure}
\plotfiddle{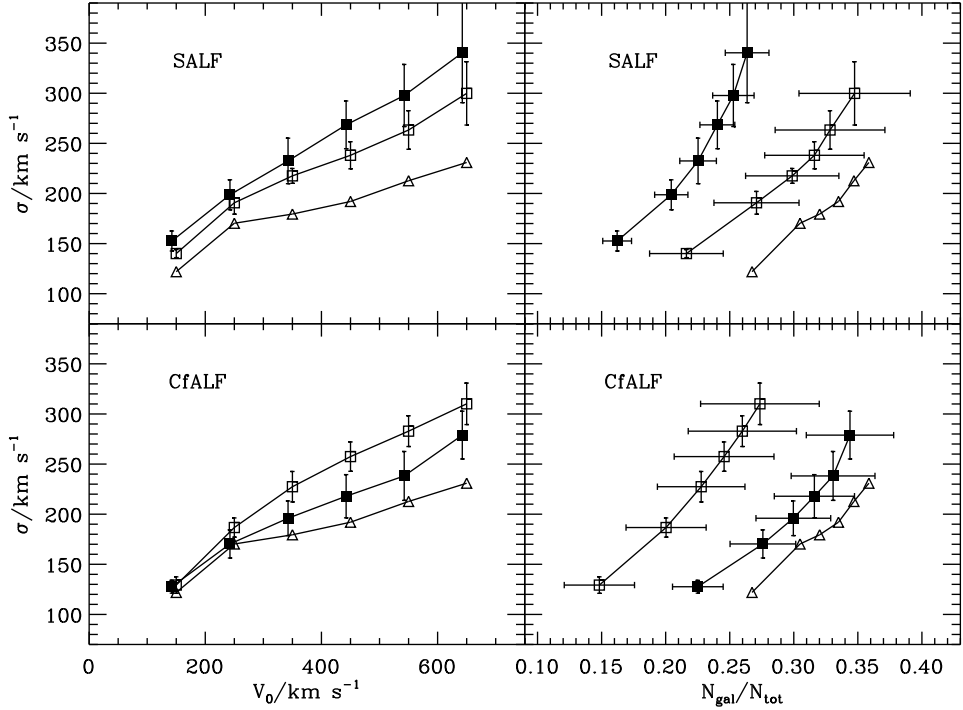}
           {0.4\vsize}              
           {90}                
           {60}                 
           {60}                 
           {210}               
           {-40}                
\caption{{\bf Left Panels}: Weighted median velocity dispersion $\sigma$ of group
catalogues versus the linking parameter $V_0$ at fixed $\delta n/n=80$. Triangles, open and solid
squares indicate the CfA2N, $\tau$CDM and $\Lambda$CDM catalogues
respectively. Model groups are extracted from
the galaxy catalogues with the semi-analytic model luminosity function (top panel) or
the imposed CfA luminosity function (bottom panel). Points show the mean values averaged
over the ensamble of ten mock catalogues. Error bars are the 1-$\sigma$ standard deviations. 
Solid squares ($\Lambda$CDM model) are slightly shifted to the left for clarity.
{\bf Right Panels}: $\sigma$ versus the fraction 
of galaxies in groups $N_{\rm gal}/N_{\rm tot}$ for the
same catalogues of the left panels. Points from left to right correspond
to increasing $V_0$. Symbols are as for the left panels.}
\label{fig:sigma_v0}
\end{figure}

The left panels of Fig. \ref{fig:sigma_v0} show that the trend of
$\sigma$ with $V_0$ is similar in all our simulated redshift surveys
and appears to be independent both of cosmological model ($\Lambda$CDM
or $\tau$CDM) and of the adopted luminosity function (SALF or CfALF).
The velocity dispersions of the simulated groups are systematically
larger than the observed values by an amount which varies from about
10\% at low $V_0$ to about 50\% at large $V_0$. This difference is
seen in all ten simulated surveys of each model. 

Let us define the grouped fraction $N_{\rm gal}/N_{\rm tot}$ as the
number of galaxies in groups divided by the total number of galaxies in the catalogue.
Nolthenius et al. (1997) 
suggest using the dependence of the pair 
($N_{\rm gal}/N_{\rm tot}$, $\sigma$) on $V_0$ at fixed $\delta n/n$ as a
diagnostic to discriminate between models. Differences in the small-scale
dynamics can show up as a shifting of the tracks in the $N_{\rm
gal}/N_{\rm tot}$ -- $\sigma$ plane. The right panels of Fig. \ref{fig:sigma_v0}
show such tracks for the CfA2N survey and for our simulated redshift surveys.
Clearly, the choice of the luminosity function affects $N_{\rm gal}/N_{\rm tot}$ strongly
but in opposite directions for our two models;
if the SALF  is used, the $\tau$CDM catalogue is closer than $\Lambda$CDM 
to the CfA2N; with the CfALF, we obtain the opposite result.
Our models yield a fraction of galaxies in groups
systematically lower  than the CfA2N sample.
When $V_0=350$ km s$^{-1}$, the fraction of galaxies in groups 
can be from $\sim 7\%$ (CfALF $\Lambda$CDM) to $\sim 30\%$ (SALF $\Lambda$CDM) smaller
than in the CfA2N sample. Note however that the CfA2N track is less than 
two standard deviations from the CfALF $\Lambda$CDM track.

\subsection{Group Abundances}{\label{sec:real_gr}}
 
\begin{figure}
\plotfiddle{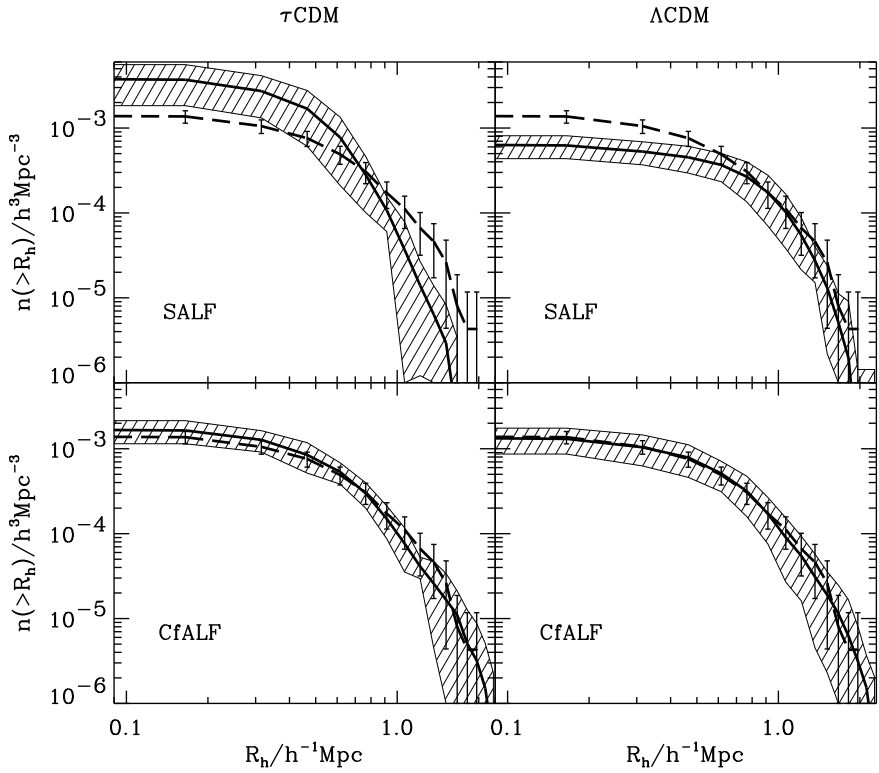}
           {0.4\vsize}              
           {90}                
           {60}                 
           {60}                 
           {210}               
           {-40}                
\caption{Group abundance by harmonic radius $R_h$.
Number densities are estimated for catalogues extracted with
$\delta n/n=80$ and $V_0=350$ km s$^{-1}$. Bold lines are the mean
number densities averaged over the ensamble of ten mock catalogues. Shaded areas show
the 3-$\sigma$ deviations. Dashed lines are for the CfA2N groups. 
Error bars on the CfA2N curves are Poisson 3-$\sigma$ deviations.}
\label{fig:gr_nd_rh}
\end{figure}

\begin{figure}
\plotfiddle{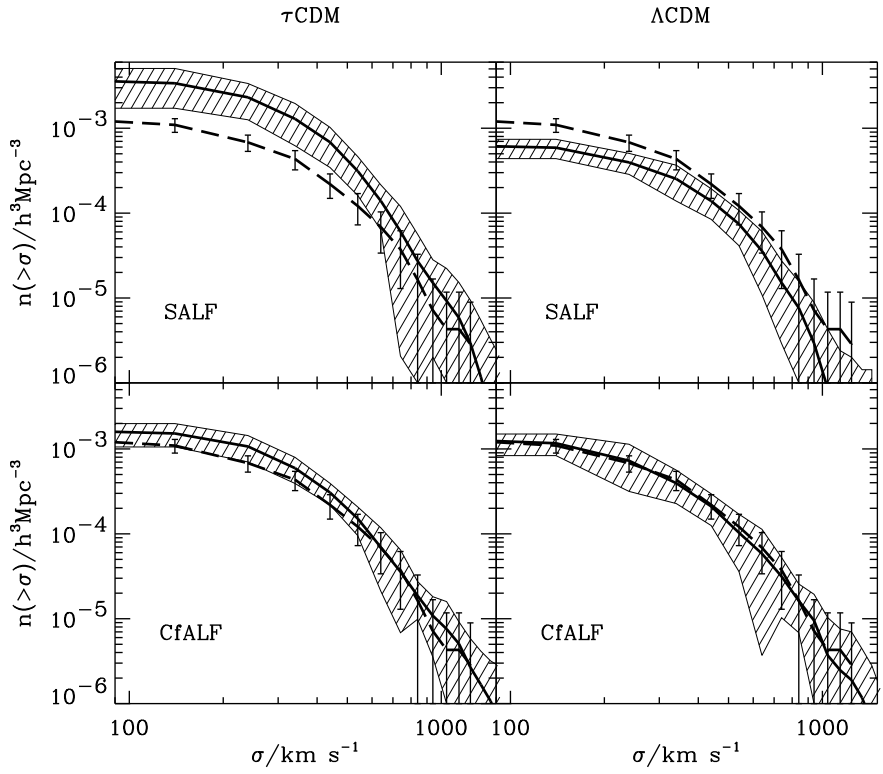}
           {0.4\vsize}              
           {90}                
           {60}                 
           {60}                 
           {210}               
           {-40}                
\caption{Group abundance by velocity dispersion $\sigma$.
Lines and deviations are as in Fig. \ref{fig:gr_nd_rh}.}
\label{fig:gr_nd_sigma}
\end{figure}

Figs. \ref{fig:gr_nd_rh} and \ref{fig:gr_nd_sigma} show 
the abundance of groups as a function of harmonic radius and velocity dispersion
derived from our various catalogues.                          
The top panels show results for the semi-analytic luminosity function (SALF). The 
discrepancy with the observations is simply a
reflection of
the overestimate (underestimate) of the total luminosity
density of the Universe by  $\tau$CDM ($\Lambda$CDM) as compared with CfA2N.
When we impose the CfA survey luminosity function (bottom panels),
the  simulation results are in much better agreement with the observations, 
especially for $\Lambda$CDM. 

The change in luminosity function from SALF to CfALF has a strong effect on the normalisation
of the group abundance function, but rather little effect on its shape. As a result, the median
properties of groups do not depend very strongly on the luminosity function.
Tables \ref{tab:gr_LF_tcdm} and \ref{tab:gr_LF_lcdm} 
list the weighted quartiles of galaxy properties in the SALF and CfALF catalogues
and compare them  with the CfA2N values. 
Taking the variations between different mock catalogues into account, the median size
and velocity  dispersion of groups in the $\Lambda$CDM simulation agree reasonably  well
with the observations, if the CfALF is adopted. 
Moreover, we note that, according to our prescription for identifying galaxies in the simulations,
the velocity of  the central galaxy coincides 
with that of the dark matter particle with the greatest absolute value of the 
gravitational potential energy. This
particle can move rapidly whereas the central galaxy is plausibly almost at
rest with respect to the barycentre of its dark halo.
When we replace the central galaxy velocity with 
the mean velocity of the dark particles within $r_{200}$, the median velocity dispersion
of groups becomes $\sim 10\%$ smaller. This effect is sufficient to bring the CfALF $\Lambda$CDM
model in good agreement with the CfA2N.
On the other hand, groups in the $\tau$CDM simulation still have 
velocity dispersions slightly higher than observed.

Note that, on average, the difference between the central galaxy velocity
and the barycentric velocity of its dark halo is $\sim 80$ km s$^{-1}$.
Replacing the central galaxy velocity by the barycentric motion has a measurable effect 
on group dispersions because most groups are triplets and quadruplets.
The same replacement
does not affect the galaxy pairwise velocity statistics
computed in section \ref{sec:csi}.

For both $\tau$CDM and $\Lambda$CDM,
groups are significantly less luminous than in the real Universe.
This is a consequence of the presence of the ``Great Wall'' in the 
CfA2N region: most of the groups lie within this structure at $\sim 7000-10000$ km s$^{-1}$
and so have a median redshift $\sim 8000$ km s$^{-1}$. In the models, however, the median
group redshift is $\sim 6000$
km s$^{-1}$, leading to a typical luminosity a factor $\sim 2$ below that found
for the CfA2N.

Note also that although for the SALF catalogues the interloper-bias
conspiracy results in estimates of $\Omega_0$ which are quite close to
the true values, the same conspiracy does not hold for the
CfALF. Indeed, when our two simulations are forced to have the same
luminosity function they produce group catalogues which are very
similar in most of their properties. In particular, they have similar
estimated mass-to-light ratios and so lead to similar estimates of
$\Omega_0$ despite the factor of 3 difference between the true density values.
  

\section{CORRELATION FUNCTIONS}\label{sec:csi}

To probe dynamics of galaxies on nonlinear scales, knowledge
of galaxy peculiar velocities is necessary. Even for nearby galaxies direct
measurements have
$\sim 20$\% uncertainties and only a relatively small number of measurements are currently
available (see e.g. \cite{Strauss95} and references therein).
Alternatively, we can study peculiar velocities statistically 
by using galaxy-galaxy redshift space correlation functions.
Marzke et al. (1995) 
compute such correlation functions for the CfA redshift surveys. 
Here, we analyse our mock catalogues following their procedures which we summarise in the
next section.  

\subsection{Method}\label{sec:xi_meth}

In redshift space, the vector ${\bf s}_i=cz_i{\bf r}_i$ locates a galaxy with redshift $cz_i\ll c$ 
and celestial coordinates ${\bf r}_i=(\alpha_i,\delta_i)$. For small angular separations 
($<50^{\rm o}$ in our analysis),
we define the components of the relative separation ${\bf s}={\bf s}_i-{\bf s}_j$ of a pair
of galaxies
\begin{equation} 
\pi={{\bf s}\cdot {\bf l}\over\vert {\bf l}\vert}, \quad r_p^2=s^2-\pi^2
\end{equation}
where ${\bf l}=({\bf s}_i+{\bf s}_j)/2$. The two-dimensional redshift space correlation
function $\xi(r_p,\pi)$ measures 
the excess probability, compared to a Poisson distribution, 
that a galaxy pair has separation $(r_p,\pi)$.
We estimate $\xi(r_p,\pi)$ by weighting each galaxy according to 
the minimum variance estimator (\cite{Davis82}).

By inverting $w(r_p)$, the projection of  $\xi(r_p,\pi)$ onto the $r_p$ axis,
we can estimate the spatial correlation function $\xi(r)$.
Assuming a power-law, $w(r_p)=Ar_p^{1-\gamma}$, gives 
$\xi(r)=(r/r_0)^{-\gamma}$, and $A=r_0^\gamma \Gamma(1/2)\Gamma[(\gamma-1)/2]/ \Gamma(\gamma/2)$.

On non-linear scales, we can then model $\xi(r_p,\pi)$ as (e.g. \cite{Fisher95}) 
\begin{equation}
1+\xi(r_p,\pi)=\int_{-\infty}^{\infty}\{1+\xi[(r_p^2+y^2)^{1/2}]\}f(r_p,\pi,y)dy
\label{eq:csi_conv}
\end{equation}
where the pairwise velocity distribution $f(r_p,\pi,y)$ is well approximated by the 
exponential form 
\begin{equation}
f(r_p,\pi,y) = C(\sigma_{12}) \exp\left[-{\sqrt{2}\over\sigma_{12}}\left\vert\pi-y
-{y\over r}\langle v_{12}(r)\rangle\right\vert\right] 
\label{eq:pairwise}
\end{equation}
where $r^2=r_p^2+y^2$, and $y$ is the pair separation in real space along the line
of sight. We model the mean streaming velocity $\langle v_{12}(r)\rangle $ with the scale-invariant
solution to the truncated BBGKY hierarchy derived by Davis \& Peebles (1983), 
\begin{equation}
\langle v_{12}(r)\rangle = -{Fr\over 1+(r/r_0)^2},
\label{eq:v12}
\end{equation}
where the constant $F=1$ in the similarity solution.

Eq. (\ref{eq:pairwise}) is an excellent approximation to 
the real distribution on very non-linear scales as already discussed by \cite{Sheth96}
and \cite{Diaferio96}. Eq. (\ref{eq:pairwise}) also describes both dark matter
particles and galaxies in our models quite accurately (\cite{Diaferio99}). 
On mildly non-linear scales, infall skews the distributions. This skewness can be 
elegantly formalized with an Eulerian perturbative approach (\cite{Juszk98}).

On the contrary, Eq. (\ref{eq:v12}) is not a good approximation to the mean
streaming velocity measured in the simulations. 
In Sect. \ref{sec:xi_res}  we will see that the uncertainty on
$\langle v_{12}(r)\rangle$ does not strongly affect the estimate 
of the pairwise velocity dispersion $\sigma_{12}$, provided  it is not assumed
to be zero on all scales.

In order to estimate the pairwise velocity dispersion $\sigma_{12}(r_p)$ 
at fixed projected separation $r_p$, 
we first determine the spatial correlation function $\xi(r)$; 
we then perform the integral in equation (\ref{eq:csi_conv})
by assuming $\sigma_{12}$ constant.
Note, however, that $\sigma_{12}$ does depend on $y=(r^2-r_p^2)^{1/2}$
because the cosmic virial theorem predicts $\sigma_{12}\propto r^{2-\gamma}$.
However, since $\gamma\approx 2$, 
this procedure yields values of $\sigma_{12}$ in reasonable agreement
with those computed with the full three-dimensional information.
Finally, we use the measured $\xi(r_p,\pi)$ to impose the normalisation $C(\sigma_{12})$ in 
equation (\ref{eq:pairwise}).
This procedure decreases the number of degrees of freedom by one.

To estimate errors on $\xi(r_p,\pi)$ we use a bootstrap procedure with 50 resampled samples.
Estimates of $\xi(r_p,\pi)$ at different separations are correlated. Therefore, 
determining $\sigma_{12}$ by minimizing the $\chi^2$ computed directly with eq. (\ref{eq:csi_conv})
is not the correct procedure. Principal component analysis (\cite{Kendall75}, \cite{Fisher94a}) 
transforms a set of correlated quantities $\{\xi\}$ into a set of uncorrelated quantities $\{x\}$. 
Meaningful results
come from the standard $\chi^2$ analysis applied to this latter set.
Note that the errors of both the original and the transformed data sets are not Gaussian distributed.
Therefore, the confidence levels of the computed $\chi^2$ should be considered only indicative.

\subsection{Results}\label{sec:xi_res}

\begin{figure}
\plotfiddle{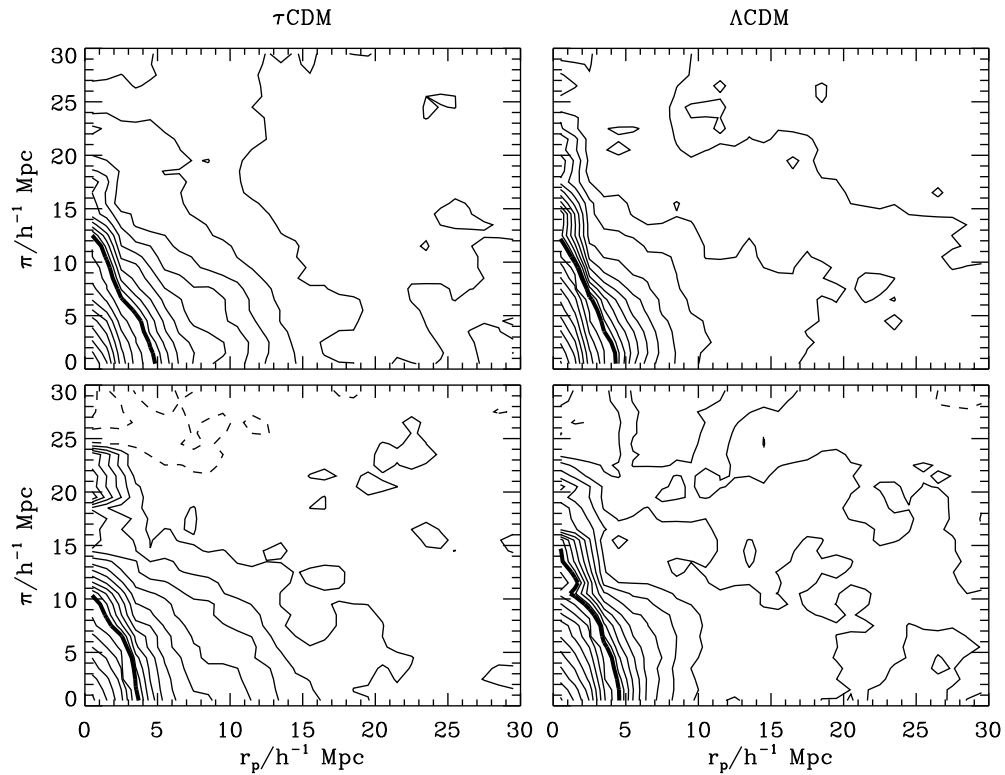}
           {0.4\vsize}              
           {90}                
           {60}                 
           {60}                 
           {210}               
           {-40}                
\caption{Maps of the two dimensional correlation function $\xi(r_p,\pi)$. 
Top panels are for the SALF catalogues of Figs. \ref{fig:slice_tcdm} and \ref{fig:slice_lcdm}. 
Bottom panels are for the corresponding CfALF catalogues. 
The bold contour indicates $\xi(r_p,\pi)=1$. Contour
levels are separated by logarithmic intervals of 0.1 for $\xi(r_p,\pi)>1$ and
by linear intervals of 0.1 for $\xi(r_p,\pi)<1$. Dashed contours indicate $\xi(r_p,\pi)<0$.}
\label{fig:csi_map}
\end{figure}

Fig. \ref{fig:csi_map} shows the $\xi(r_p,\pi)$ maps of our SALF and
CfALF catalogues. At  $r_p\lesssim 5 h^{-1}$ Mpc, 
$\xi(r_p,\pi)$ in the SALF $\tau$CDM catalogue is larger
than in the CfALF catalogue; the SALF catalogue contains more galaxies in groups and
clusters. On the other hand 
the two $\xi(r_p,\pi)$ maps for the $\Lambda$CDM model are very similar,
perhaps because the differences between the two luminosity functions are smaller.               

We can quantify the differing behaviour of our two models by looking at 
the parameters $r_0$ and $\gamma$ of the real space correlation 
function derived by fitting a power law to the projected correlation function 
$w(r_p)$. 
Table \ref{tab:xi} shows that
$r_0$ decreases by $35\%$ from the SALF
to the CfALF catalogue in the $\tau$CDM model, while in 
the $\Lambda$CDM model there is no significant change. 
None of these models matches the CfA2N parameters satisfactorily, perhaps because
the CfA2N yields a correlation length  which is significantly 
larger than in 
other surveys: CfA2 South and SSRS2 have $r_0=4.75$ and $5.08 h^{-1}$ Mpc respectively
(\cite{Marzke95}), while the LCRS gives $5.06 h^{-1}$ Mpc (\cite{Jing98}); 
these values are close to those of our models.
Note that CfA2 South also has $\gamma=1.99$, so that its correlation 
function is in close agreement with that of our $\Lambda$CDM model. 

\begin{figure}
\plotfiddle{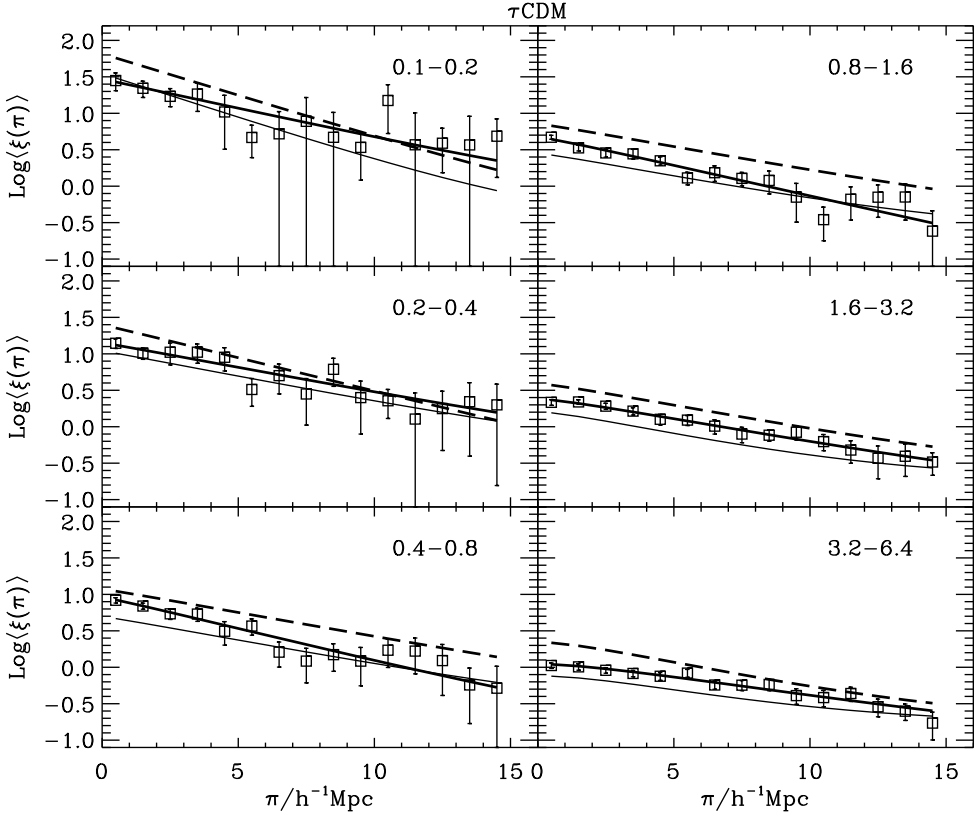}
           {0.4\vsize}              
           {90}                
           {60}                 
           {60}                 
           {210}               
           {-40}                
\caption{Correlation functions projected onto the $\pi$ axis for different
intervals of the projected separation $r_p$ for the SALF $\tau$CDM catalogue shown
in Fig. \ref{fig:slice_tcdm} (squares).
Intervals of $r_p$ in units of $h^{-1}$ Mpc
are shown in the right upper corner of each panel. Bold solid lines show
the best fits to eq. (\ref{eq:csi_aver}). Thin solid lines 
are the best fits for the corresponding CfALF catalogue. Long dashed lines are
the best fits for the CfA2N catalogue. All curves are for $F=1$ in eq. (\ref{eq:v12}).
$\langle \xi(\pi)\rangle$ curves for the ten mock catalogues can differ
by $\sim 10-20\%$ depending on the $r_p$ interval.}
\label{fig:xi_tcdm}
\end{figure}

\begin{figure}
\plotfiddle{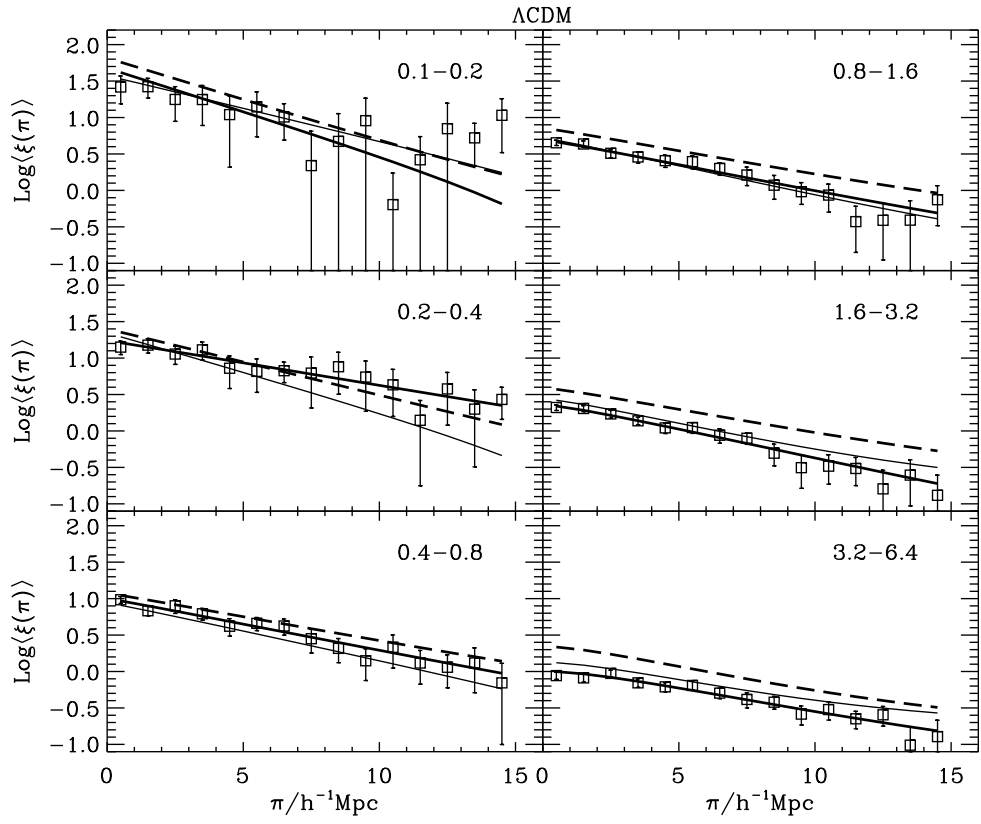}
           {0.4\vsize}              
           {90}                
           {60}                 
           {60}                 
           {210}               
           {-40}                
\caption{Same as Fig. \ref{fig:xi_tcdm} for the $\Lambda$CDM catalogue of
Fig. \ref{fig:slice_lcdm}.}
\label{fig:xi_lcdm}
\end{figure}

We find a similar result by looking 
at the projection of $\xi(r_p,\pi)$ 
onto the $\pi$ axis
averaged over the $r_p$ interval $(r_p^1,r_p^2)$ 
\begin{equation}
\langle \xi(\pi)\rangle={1\over r_p^2-r_p^1} \int_{r_p^1}^{r_p^2} \xi(r_p,\pi) dr_p
\label{eq:csi_aver}
\end{equation}
(Figs. \ref{fig:xi_tcdm} and \ref{fig:xi_lcdm}). 
Independently of the luminosity function, both models predict a correlation amplitude 
somewhat smaller than that of the CfA2N.
Typical amplitude ratios between the CfA2N and the models are in the range $\sim 1.0-2.3$. 
These differences are similar to those between 
the CfA2N and CfA2 South or SSRS2 (\cite{Marzke95}).

\begin{figure}
\plotfiddle{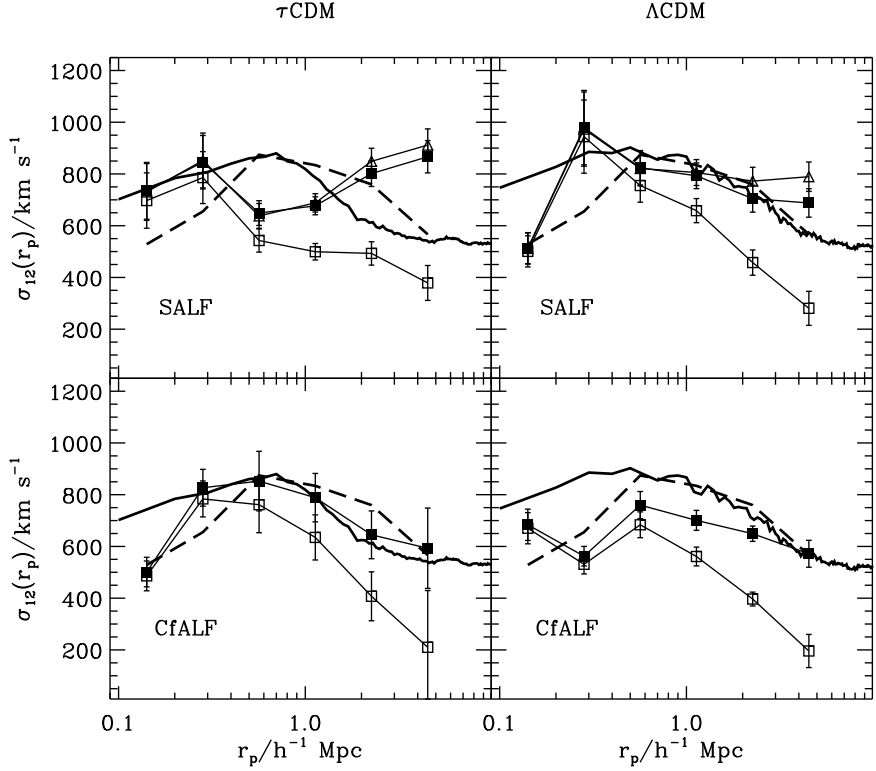}
           {0.4\vsize}              
           {90}                
           {60}                 
           {60}                 
           {210}               
           {-40}                
\caption{Pairwise velocity dispersion $\sigma_{12}$ at different projected separations
$r_p$ corresponding to the curves in Figs. \ref{fig:xi_tcdm} and \ref{fig:xi_lcdm}. 
Solid (open) squares are for $F=1$ ($F=0$) in eq. (\ref{eq:v12}). In the top panels, 
triangles are for $F=2.5$ and $r_0=3 h^{-1}$ Mpc in eq. (\ref{eq:v12}).
The $\sigma_{12}(r_p)$ profiles have a scatter of $\sim 15\%$ for different
mock catalogues. 
Bold lines show the three-dimensional pairwise velocity
dispersion of galaxies brighter than $M_B=-17.5+5{\rm Log}h$ (Paper I); the dashed line is
the CfA2N profile when $F=1$ (\cite{Marzke95}).} 
\label{fig:combine_3sigma}
\end{figure}

Small variations in $\langle \xi(\pi)\rangle$ translate into large
variations in the best fit parameters $\sigma_{12}$.
In fact, Marzke et al. (1995) 
show that $\sigma_{12}$ can vary by a factor
as large as three from survey to survey.
In Fig. \ref{fig:combine_3sigma},
solid squares show $\sigma_{12}(r_p)$ when the similarity solution 
for the mean streaming velocity is assumed ($F=1$ in eq. [\ref{eq:v12}]).
The bold solid line is the
$\sigma_{12}$ of galaxies computed with the full three-dimensional information. 
The variations in the amplitude and slope of $\langle \xi(\pi)\rangle$ 
between SALF and CfALF catalogues
make $\sigma_{12}$ vary by $\sim 200$ km s$^{-1}$. 
However, given the sensitivity of $\sigma_{12}$, the differences between
the models are not significant: 
both models predict a $\sigma_{12}(r_p)$ profile close to that of the CfA2N, 
independently of the luminosity function adopted. 
A more robust statistic, e.g. the
single-particle-weighted statistic $\sigma_1$ suggested by \cite{Davis97},
might provide a more convincing comparison. 

Finally, we note that in our models $\langle v_{12}(r)\rangle$ is significantly different from zero
at megaparsec scales; 
moreover, the similarity solution (eq. [\ref{eq:v12}]) does not describe its
behaviour adequately. However, the estimated $\sigma_{12}$ are only
weakly dependent on the form of $\langle v_{12}(r)\rangle$. For comparison, 
we estimate $\sigma_{12}$ with $F=2.5$ and $r_0=3h^{-1}$ Mpc 
which appears to  be a better fit to $\langle v_{12}(r)\rangle$ in our simulations.
The agreement between the estimated $\sigma_{12}(r_p)$ and the three-dimensional profile does not
improve substantially (open triangles in the top panels of Fig. \ref{fig:combine_3sigma}).
On the other hand, in agreement with the result of Marzke et al. (1995), 
assuming $F=0$ underestimates $\sigma_{12}(r_p)$ significantly (open squares in 
Fig. \ref{fig:combine_3sigma}). This raises a serious
question about the Fourier method suggested by \cite{Landy98} for estimating
$\sigma_{12}(r_p)$: this method necessarily assumes $\langle v_{12}(r)\rangle=0$.





\section{CONCLUSION}\label{sec:disc}

In Paper I, we simulated the formation, evolution and clustering 
of galaxies by combining dissipationless $N$-body simulations and semi-analytic
models of galaxy formation. We investigated a high density universe ($\tau$CDM) and
a low density universe with a cosmological constant ($\Lambda$CDM). Here, we 
extract wide angle mock redshift surveys from these simulations and compare them
with the CfA2N redshift survey by compiling catalogues
of galaxy groups and by computing the redshift space correlation function $\xi(r_p,\pi)$.

Despite their very different $\Omega_0$ values, both
models yield a reasonable match to the data, although both 
appear  slightly less clustered than the CfA2N: for example, 
$\sim 30\%$ ($\tau$CDM) and $\sim 23\%$ ($\Lambda$CDM) of galaxies
are in groups rather than the $32\%$ observed; moreover, the amplitude
of $\xi(r_p,\pi)$ is systematically a factor $\sim 1.5-2.0$ smaller than
observed. However, this weaker clustering 
is similar to that observed in the southern regions of the CfA redshift surveys and in the LCRS, 
suggesting that galaxies within the CfA2N catalogue might be unusually highly clustered.

Our semi-analytic modelling predicts luminosity functions which 
fit the CfA results rather poorly. To test whether
this disagreement induces clustering differences between our mock surveys and the
CfA2N galaxy catalogue, we impose the CfA survey luminosity function on the models. 
We find that in fact only the group abundances are strongly affected, 
becoming substantially closer to those observed in the modified catalogues. Quantities
connected with the underlying mass distribution, such as the median velocity dispersion
of groups or the redshift space correlation function, 
show less sensitivity to the adopted luminosity function. 

We compare our models with the CfA2N because (1) this survey 
provides the largest catalogue of galaxy groups currently available, and (2) it 
covers a volume similar to that we have simulated. However,
a coherent two-dimensional structure, the ``Great Wall'', dominates the
large scale distribution of galaxies in the CfA2N and is not
reproduced in our simulations. This structure plays a significant role in determining 
the typical luminosities of the observed groups and may also affect other
group properties. It is unclear whether the apparent deficiency of
 coherent two-dimensional structure in our mock catalogues 
is a significant problem for our cosmological models, 
or reflects cosmic variance and the relatively small size of our simulation boxes.  
The modest discrepancies between
our models and the CfA2N could plausibly disappear when other surveys,
covering larger volumes and based on better photometric data
become available, and can be compared with simulations of similarly large cosmological volumes.

Given the limitations of the observational and numerical data we consider in this paper,
the agreement between theory and observation is remarkable and demonstrates that  
present day group dynamics alone give little information about $\Omega_0$. Our preference for
$\Lambda$CDM over $\tau$CDM is based on its better agreement with
the observed luminosity functions and Tully-Fisher relations (Paper I) and with the
observed evolution of clustering (Paper II) rather than on any difference in pairwise
velocities or in the mass-to-light values implied for groups.
On the other hand, these clustering differences reflect observable differences in the evolution
of groups and clusters and of their galaxy populations which should be detectable 
in future surveys to higher redshift. We will investigate
these issues more thoroughly in a forthcoming paper.

\acknowledgements
{\bf Acknowledgements}

The $N$-body simulations were carried
out at the Computer Center of the Max-Planck Society
in Garching and at the EPPC in Edinburgh, as part of the Virgo Consortium project.
We thank Margaret Geller, John Huchra and Ron Marzke for enabling us
to compare the CfA2N data directly with these numerical data.
We also thank Margaret Geller, Massimo Ramella and Roberto Trasarti-Battistoni
for several enlightening discussions and an anonymous referee for suggesting
improvements in the presentation of our results.
During this project, A.D. was a Marie Curie Fellow and 
held grant ERBFMBICT-960695 of the Training and Mobility of
Researchers program financed by the European Community.
A.D. also acknowledges support from an MPA guest post-doctoral fellowship.

\appendix
\section{APPENDIX}

For reference, we review here the formulae we use to compute the physical properties 
of groups.

A system with $N$ galaxies has three-dimensional (3D) harmonic radius 
\begin{equation}
R_h = {N(N-1)\over 2} \left(\sum_{i=1}^{N-1}\sum_{j=i+1}^N 
{1\over \vert{\bf r}_{ij}\vert} \right)^{-1}
\end{equation}
where ${\bf r}_{ij}$ are the pairwise galaxy separations.
The corresponding harmonic radius for a real system at redshift 
$\langle cz\rangle$ is 
\begin{equation}
R_h = {\pi\over 2} {\langle cz\rangle\over H_0} N(N-1) \left[\sum_{i=1}^{N-1}\sum_{j=i+1}^N
{1\over \tan(\theta_{ij}/2)}\right]^{-1}
\end{equation}
where $\theta_{ij}$ are the pairwise galaxy angular separations and $H_0$ is the
Hubble constant.
 
The one-dimensional velocity dispersion from the 3D velocities is
\begin{equation}
\sigma = \left[{1\over 3(N-1)}\sum_{i=1}^N({\bf v}_i-\langle {\bf v}\rangle)^2\right]^{1/2},
\end{equation}
whereas from the line-of-sight velocities it is
\begin{equation}
\sigma = \left[{1\over N-1}\sum_{i=1}^N(cz_i-\langle cz\rangle)^2\right]^{1/2}.
\end{equation}

The combination of $R_h$ and $\sigma$ provides the virial mass via the virial theorem
\begin{equation}
M_{\rm vir} = {6 \sigma^2 R_h\over G}
\end{equation}
where $G$ is the gravitational constant.

The total luminosity of a galaxy group is the sum of the contribution from the $N$
galaxies observed and the contribution from galaxies which are too faint
to be observed:
\begin{equation}
L_{\rm tot} = L_{\rm faint} + \sum_{i=1}^N L_i.
\end{equation}
We can estimate the contribution of the faint galaxies by assuming the
luminosity function to be universal:
\begin{equation}
L_{\rm faint} = {\rm Volume}  \times \langle L_{\rm faint}\rangle  =
{N\over \int_{L_{\rm lim}}^\infty \phi(L)dL} \times \int_0^{L_{\rm lim}} L\phi(L)dL
\end{equation}
where $L_{\rm lim}$ is the luminosity of the faintest observable galaxy within the
group.
We adopt the usual Schechter form of the luminosity function
\begin{equation}
\phi(L)dL=\phi^*(L/L^*)^\alpha \exp(-L/L^*) d(L/L^*)
\end{equation}
or equivalently
\begin{equation}
\phi(M)dM= 0.4\ln 10 \phi^* 10^{-0.4(M-M^*)(\alpha+1)} \exp[-10^{-0.4(M-M^*)}]dM
\end{equation}
when absolute magnitudes $M$ are used.

\clearpage
\begin{deluxetable}{lccc}
\tablecaption{Luminosity Function Parameters}
\tablehead{ \colhead{ } & \colhead{$\alpha$} & \colhead{$M_B^*-5{\rm Log}h$} &
\colhead{$\phi^*/{\rm gal}\ {\rm mag}^{-1} h^3 {\rm Mpc}^{-3}$} }
\startdata
$\tau$CDM & $-1.67\pm0.01$ & $-21.05\pm 0.03$ & $(5.8\pm1.0)\times10^{-3}$ \nl
$\Lambda$CDM & $-1.40\pm0.01$ & $-20.13\pm0.02$ & $(4.1\pm0.8)\times10^{-3}$   \nl
CfA & $-1.00\pm0.20$ & $-18.80\pm0.30$ & $(4.0\pm1.0)\times10^{-2}$   \nl
\enddata
\tablecomments{Parameters and 1-$\sigma$ standard deviations
of the Schechter function fit to the model luminosity
functions (Fig. \ref{fig:lum_fun_cl}) and the CfA Redshift Survey (\cite{MarzkeLF94}).}
\label{tab:lum_fun}
\end{deluxetable}

\begin{deluxetable}{lcc}
\tablecaption{3D Groups}
\tablehead{ \colhead{ } &
\colhead{$\tau$CDM} & \colhead{$\Lambda$CDM} }
\startdata
$N_{\rm gal}/N_{\rm tot}$ & 0.30 &  0.25 \nl
$R_h$ & 0.13/0.21/0.29 & 0.15/0.27/0.43 \nl
$\sigma$&  149/213/296  & 179/246/335 \nl
${\rm Log}M_{\rm vir}$  & 12.69/13.05/13.47 & 12.92/13.30/13.69 \nl
$M_{\rm vir}/M_{200}$ & 0.88/1.37/2.04 & 0.76/1.33/2.13 \nl
${\rm Log}L_B$ &  10.53/10.66/10.89 & 10.48/10.61/10.83 \nl
${\rm Log}(M_{\rm vir}/L_B)$ & 2.09/2.37/2.61 & 2.39/2.67/2.93 \nl
$\Omega_0^{\rm est}$ & 0.55 &  0.18\nl
\enddata
\tablecomments{Quartiles of the distributions for the galaxy groups
within the box.  $R_h$, $\sigma$, $M_{\rm vir}$, $L_B$, and
$M_{\rm vir}/L_B$ are in units of $h^{-1}$ Mpc, km s$^{-1}$, $h^{-1}M_\odot$, $h^{-2}L_\odot$, and
$hM_\odot/L_\odot$, respectively. $M_{200}$ is the dark halo mass within $r_{200}$.}
\label{tab:gr_3D}
\end{deluxetable}

\begin{deluxetable}{lccc}
\tablecaption{Redshift Space Groups: $\tau$CDM}
\tablehead{ \colhead{ } &
\colhead{RS} &  \colhead{RS$_{\rm c}$} & \colhead{3D} }
\startdata
$N_{\rm gal}/N_{\rm tot}$ & 0.30 & 0.20 & 0.30 \nl
$R_h$ & 0.22/0.35/0.55 &  0.15/0.25/0.32 & 0.13/0.21/0.29 \nl
$\sigma$ & 121/216/336 &  151/227/378 & 149/213/296 \nl
${\rm Log}M_{\rm vir}$ & 12.87/13.25/13.79 & 12.73/13.15/13.80 & 12.69/13.05/13.47 \nl
${\rm Log}L_B$ & 10.51/10.69/10.93 &  10.61/10.77/10.99 & 10.53/10.66/10.89 \nl
${\rm Log}(M_{\rm vir}/L_B)$ & 2.14/2.49/2.95 & 1.92/2.36/2.71 & 2.09/2.37/2.61 \nl
$\Omega_0^{\rm est}$ & 0.72 &  0.54 & 0.55 \nl
\enddata
\tablecomments{Weighted quartiles of the distributions for the galaxy group catalogue from
the $\tau$CDM SALF survey shown in Fig. \ref{fig:slice_tcdm}: 
RS refers to groups identified in redshift space, RS$_{\rm c}$ 
to groups identified in redshift space with interlopers excluded, and 3D
to properties derived from the 3D galaxy distribution.
Quantities are in the same units as in Table \ref{tab:gr_3D}.}
\label{tab:gr_RS_tcdm}
\end{deluxetable}

\begin{deluxetable}{lccc}
\tablecaption{Redshift Space Groups: $\Lambda$CDM}
\tablehead{ \colhead{ } &
\colhead{RS} &  \colhead{RS$_{\rm c}$} & \colhead{3D} }
\startdata
$N_{\rm gal}/N_{\rm tot}$ & 0.23 & 0.15 & 0.25 \nl
$R_h$ & 0.35/0.59/0.80 &  0.18/0.34/0.55 & 0.15/0.27/0.43 \nl
$\sigma$ & 135/222/345 &  164/223/368 & 179/246/335 \nl
${\rm Log}M_{\rm vir}$ & 12.97/13.53/13.91 & 12.88/13.37/13.80 & 12.92/13.30/13.69 \nl
${\rm Log}L_B$ & 10.44/10.65/10.90 &  10.54/10.71/11.00 & 10.48/10.61/10.83 \nl
${\rm Log}(M_{\rm vir}/L_B)$ & 2.37/2.82/3.22 & 2.28/2.64/3.00 & 2.39/2.67/2.93 \nl
$\Omega_0^{\rm est}$ & 0.25 &  0.17 & 0.18 \nl
\enddata
\tablecomments{Same as Table \ref{tab:gr_RS_tcdm} for the $\Lambda$CDM 
SALF survey shown in Fig. \ref{fig:slice_lcdm}.}
\label{tab:gr_RS_lcdm}
\end{deluxetable}

\begin{deluxetable}{lccc}
\tablecaption{Luminosity Function Effect: $\tau$CDM}
\tablehead{ \colhead{ } &
\colhead{SALF} & \colhead{CfALF} & \colhead{CfA2N}  }
\startdata
$N_{\rm gal}/N_{\rm tot}$ & 0.30 & 0.22 & 0.32 \nl
$R_h$ & 0.22/0.35/0.55 & 0.21/0.39/0.59 & 0.23/0.44/0.71 \nl 
$\sigma$ & 121/216/336 & 134/225/336 & 99/183/299 \nl 
${\rm Log}M_{\rm vir}$ & 12.87/13.25/13.79 & 12.80/13.39/13.82 & 12.60/13.19/13.84 \nl 
${\rm Log}L_B$ & 10.51/10.69/10.93 & 10.34/10.52/10.79 & 10.52/10.88/11.19 \nl 
${\rm Log}(M_{\rm vir}/L_B)$ & 2.14/2.49/2.95 & 2.28/2.76/3.14 & 1.77/2.43/2.84 \nl 
$\Omega_0^{\rm est}$ & 0.72 & 0.43 & 0.20 \nl 
\enddata
\tablecomments{Weighted quartiles of the distributions for the redshift space group catalogues
from two corresponding $\tau$CDM surveys differing in the luminosity function adopted.
The SALF survey is shown in Fig. \ref{fig:slice_tcdm}.
CfA2N column is for the groups in the CfA2N catalogue.
Quantities are in the same units as in Table \ref{tab:gr_3D}.}
\label{tab:gr_LF_tcdm}
\end{deluxetable}

\begin{deluxetable}{lccc}
\tablecaption{Luminosity Function Effect: $\Lambda$CDM}
\tablehead{ \colhead{ } &
\colhead{SALF} & \colhead{CfALF} & \colhead{CfA2N}  }
\startdata
$N_{\rm gal}/N_{\rm tot}$ & 0.23 & 0.30 & 0.32 \nl
$R_h$ & 0.35/0.59/0.80 & 0.31/0.46/0.64 & 0.23/0.44/0.71 \nl 
$\sigma$ & 135/222/345 & 122/207/305 & 99/183/299 \nl 
${\rm Log}M_{\rm vir}$ & 12.97/13.53/13.91 & 12.87/13.38/13.77 & 12.60/13.19/13.84 \nl 
${\rm Log}L_B$ & 10.44/10.65/10.90 & 10.45/10.65/10.87 & 10.52/10.88/11.19 \nl 
${\rm Log}(M_{\rm vir}/L_B)$ & 2.37/2.82/3.22 & 2.27/2.69/3.10 & 1.77/2.43/2.84 \nl 
$\Omega_0^{\rm est}$ & 0.25 & 0.37 & 0.20 \nl 
\enddata
\tablecomments{Same as Table \ref{tab:gr_LF_tcdm} for two $\Lambda$CDM surveys. The
SALF survey is shown in Fig. \ref{fig:slice_lcdm}.}
\label{tab:gr_LF_lcdm}
\end{deluxetable}

\begin{deluxetable}{lccc}
\tablecaption{Correlation Function $\xi(r)$}
\tablehead{ \colhead{ } &
\colhead{$r_0/h^{-1}$ Mpc} & \colhead{$\gamma$} & \colhead{$\chi_{18}^2$}  }
\startdata
$\tau$CDM SALF & $5.00\pm 0.13$  & $1.69\pm0.03$ & 0.69  \nl
$\tau$CDM CfALF & $3.20\pm 0.28$  & $1.95\pm0.11$ & 0.51  \nl
\nl
$\Lambda$CDM SALF & $4.45\pm 0.11$  & $2.03\pm0.03$ & 1.24  \nl
$\Lambda$CDM CfALF & $4.33\pm 0.22$  & $2.02\pm0.07$ & 1.24  \nl
\nl
CfA2N	& $5.59\pm 0.24$  & $1.82\pm0.04$ & 0.63 \nl
\enddata
\tablecomments{Parameters of the correlation function fits for individual mock
catalogues. The SALF catalogues are shown in Figs. \ref{fig:slice_tcdm} and \ref{fig:slice_lcdm}.
The scatter between values from different mock catalogues is about $5\%$.}
\label{tab:xi}
\end{deluxetable}

\clearpage

\clearpage

\clearpage

\clearpage

\clearpage

\clearpage

\clearpage

\clearpage

\clearpage

\clearpage

\clearpage

\clearpage

\clearpage

\clearpage

\clearpage

\clearpage

\clearpage

\end{document}